\documentclass[journal]{IEEEtran}

\usepackage{caption}
\usepackage{subcaption}
\usepackage{amsfonts}
\usepackage{graphicx,amsmath,amsthm,amssymb}
\usepackage{fancyhdr}
\usepackage{dsfont}
\usepackage{array,color}
\usepackage{bm}
\usepackage{float}
\usepackage{algpseudocode}
\usepackage{multirow}
\usepackage{booktabs}
\usepackage{multirow}
\usepackage{makecell}
\usepackage{colortbl}
\usepackage{graphicx}
\usepackage{epstopdf}
\usepackage{tabularx}
\usepackage{booktabs}
\usepackage{multirow}
\usepackage{makecell}

\usepackage{etoolbox}

\usepackage[colorlinks=true]{hyperref}
\hypersetup{
    citecolor = {blue},
}

\usepackage[linesnumbered, ruled, boxed]{algorithm2e}
\usepackage{stmaryrd}
\SetKwProg{Init}{Initialization}{}

\allowdisplaybreaks[4]

\definecolor{headercolor}{RGB}{52, 101, 164}
\definecolor{rowcolor1}{RGB}{242, 246, 252}
\definecolor{rowcolor2}{RGB}{255, 255, 255}

\newtheorem{theorem}{Theorem}

\newtheorem{lemma}{Lemma}

\newtheorem{proposition}{Proposition}

\newtheorem{corollary}{Corollary}

\newtheorem{property}{Property}

\newtheorem{remark}{Remark}

\newtheorem{claim}{Claim}

\newcolumntype{A}{>{\hsize=0.95\hsize\raggedright\arraybackslash}X}  %
\newcolumntype{B}{>{\hsize=0.6\hsize\centering\arraybackslash}X}  %
\newcolumntype{C}{>{\hsize=1.7\hsize\centering\arraybackslash}X}  %
\newcolumntype{D}{>{\hsize=0.75\hsize\centering\arraybackslash}X}  %
\makeatletter
\renewcommand{\maketag@@@}[1]{\hbox{\m@th\normalsize\normalfont#1}}%
\makeatother

\begin{document}

\title{{Indoor Fluid Antenna Systems} \\ {Enabled by Layout-Specific Modeling  and} \\ {Group Relative Policy Optimization}}

\author{Tong Zhang, %
            Qianren Li, 
            Shuai Wang, \textit{Senior Member, IEEE},  	 
            Wanli Ni, 
	   Jiliang Zhang, \textit{Senior Member, IEEE},\\
	   Rui Wang,
           Kai-Kit Wong, \textit{Fellow, IEEE}, and 
           Chan-Byoung Chae, \textit{Fellow, IEEE}
\vspace{-2mm}

\thanks{T. Zhang is with Guangdong Provincial Key Laboratory of Aerospace Communication and Networking Technology, Harbin Institute of Technology, Shenzhen, 518055, China (e-mail: tongzhang@hit.edu.cn).

Q. Li and R. Wang are with the Southern University of Science and Technology (e-mail: liqr2022@mail.sustech.edu.cn, wangr@sustech.edu.cn).

S. Wang is with the Shenzhen Institutes of Advanced Technology, Chinese Academy of Sciences, Shenzhen 518055, China (e-mail: s.wang@siat.ac.cn).

W. Ni is with the Department of Electronic Engineering, Tsinghua University, Beijing 100084, China (e-mail: niwanli@tsinghua.edu.cn).

J. Zhang is with The State Key Laboratory of Synthetical Automation for Process Industries and The College of Information Science and Engineering, Northeastern University, Shenyang, China, and also with The National Mobile Communications Research Laboratory, Southeast University, China (e-mail: zhangjiliang1@mail.neu.edu.cn).

K. K. Wong is with the Department of Electronic and Electrical Engineering, University College London, Torrington Place, U.K. He is also affiliated with the Yonsei Frontier Laboratory, Yonsei University, Seoul, 03722, South Korea (e-mail: kai-kit.wong@ucl.ac.uk).

C.-B. Chae is with the School of Integrated Technology, Yonsei University, Seoul 03722 South Korea (e-mail: cbchae@yonsei.ac.kr).

T. Zhang and Q. Li contributed equally.

Correspondence Authors: J. Zhang and S. Wang.
}

\thanks{Our GRPO code is available at {\color{cyan}https://github.com/QianrenLi/rt\_grpo}.}
}

\maketitle

\begin{abstract}

Fluid antenna system (FAS) revolutionizes wireless communications via utilizing position-flexible antennas that dynamically optimize channel conditions and mitigate multipath fading. This innovation is particularly valuable in indoor environments, in which signal propagation is severely degraded due to structural obstructions and complex multipath reflections. In this paper, we investigate the channel modeling and the joint optimization of antenna positioning, beamforming, and power allocation for indoor FAS. In particular, we propose a layout-specific channel model, and employ the novel group relative policy optimization (GRPO) algorithm for tackling the optimization problem. Compared to the state-of-the-art Sionna model, our model achieves an $83.3\%$ reduction in computation time with an approximately $3$ dB increase in root-mean-square error (RMSE). When simplified to a two-ray model, our model allows for a closed-form antenna position solution with near-optimal performance. For the joint optimization problem, our GRPO algorithm outperforms proximal policy optimization (PPO) and other baselines in sum-rate, while requiring only $50.8\%$ computational resources of PPO, thanks to its group advantage estimation. Simulation results show that increasing either the group size or trajectory length in GRPO does not yield significant improvements in sum-rate, suggesting that these parameters can be selected conservatively without sacrificing performance.
\end{abstract}


\vspace{-3mm}
\section{Introduction}

\begin{table*}[t]
 	\centering
 	\caption{\textsc{Comparison of This Work and Related Studies on FAS}}\label{tab:compare}
 	\label{tab:optimization_comparison}
 	\begin{tabular}{c|c|c|c|c|c|c|c|c|c}
 		\hline
 		\multicolumn{1}{c|}{\multirow{2}{*}{\textbf{Ref.}}} & 
 		\multicolumn{1}{c|}{\multirow{2}{*}{Site-Specific}} & 
 		\multicolumn{2}{c|}{\textbf{Considered Scenario}} & \multicolumn{3}{c|}{\textbf{Optimization Variables}}
 		& 
 		\multicolumn{3}{c}{\textbf{Algorithmic Design}} \\
 		\cline{3-10}
 		& & Single-User & Multi-User & FAS Position & Beamforming & Transmit Power & Numerical Methods & DRL & Closed-Form \\
 		\hline
 		\cite{New} & & $\checkmark$ &   & $\checkmark$ & $\checkmark$ & $\checkmark$  & $\checkmark$& & \\
 		\hline
 		\cite{Mei} & & $\checkmark$ & $\checkmark$ & $\checkmark$ & $\checkmark$ & $\checkmark$ & $\checkmark$ &  & \\
 		\hline
        	{\cite{Yao}} & & {$\checkmark$} &   & {$\checkmark$} &   & {$\checkmark$}  & {$\checkmark$}  &  & {$\checkmark$} \\
    \hline 
 		\cite{Cheng} & & $\checkmark$ & $\checkmark$ & $\checkmark$ & $\checkmark$ & $\checkmark$ &$\checkmark$  & & \\
 		\hline
 		\cite{Qin} & & $\checkmark$ & $\checkmark$ & $\checkmark$ & $\checkmark$ & $\checkmark$  &$\checkmark$  &  & \\
 		\hline       
        {\cite{HaoT}} & & {$\checkmark$} & {$\checkmark$} & {$\checkmark$} & {$\checkmark$} & {$\checkmark$}  & {$\checkmark$}  &  & \\
 		\hline  
         {\cite{Ye}} & & {$\checkmark$} & {$\checkmark$} & {$\checkmark$} & {$\checkmark$} & {$\checkmark$}  &{$\checkmark$}  &  & \\
 		\hline  
 		\cite{Chao} & & $\checkmark$ & $\checkmark$ & $\checkmark$ & $\checkmark$ & $\checkmark$ & & $\checkmark$ & \\
 		\hline
 		\textbf{Ours} &$\checkmark$ & $\checkmark$ & $\checkmark$ & $\checkmark$ & $\checkmark$ & $\checkmark$ & & $\checkmark$ & $\checkmark$ \\
 		\hline
 	\end{tabular}
\vspace{-3mm}
 \end{table*}

\IEEEPARstart{M}{odern} wireless communications technologies have continuously transformed, with each generation marking a significant leap forward. The ongoing evolution from the fifth-generation (5G) to the sixth-generation (6G) represents a critical juncture in wireless communications development, characterized by groundbreaking technological shifts and innovative design approaches \cite{6GKit,dang2020should,Zhang-2024}. Despite numerous technological innovations, multiple-input multiple-output (MIMO) technique stands out as the most significant breakthrough in wireless communications of our time. One of MIMO's major innovations lies in its multiplexing gain, which allows simultaneous transmission of multiple data streams over the same frequency channel by leveraging multiple antennas \cite{mimo-1}. Despite its prominent advantages, MIMO suffers from environment-depending channel, which often leads to unstable performance if the number of antennas is not sufficient. 
 
To mitigate this issue, fluid antenna system (FAS), a revolutionary advancement in antenna technology and wireless communications, was recently proposed in \cite{CL-FAS-1,FAS,FAS-Tutorial,Lu-2025}. It can be any software-controlled structures that can dynamically alter their physical configuration to optimize the radiating response according to the environments, which can be realized using a variety of reconfigurable antenna technologies such as \cite{shen2024design,Shamim-2025,zhang2024pixel,Liu-2025arxiv}. A key application of FAS is to circumvent deep fade or co-channel interference by changing the antenna position \cite{FAS-MAC}. Traditional channel models are typically assumed fixed antenna positions, whereas FAS channel models (e.g., \cite{Wong, Mohamed-Slim, Nakagami, FAS-ISAC}) must account for the impact of dynamically changing antenna positions on channel response, thus exhibiting position dependency. However, the current research on FAS primarily focuses on nonspecific scenarios or outdoors. Such FAS  channel models include Rayleigh model \cite{Wong}, Jake's model \cite{Mohamed-Slim}, Nakagami model \cite{Nakagami}, and field response model \cite{FAS-ISAC}. But it was reported by Ericsson that $70\%\sim80\%$ of mobile data traffic is generated indoors \cite{ericsson2023indoor}. Unlike outdoor scenarios, indoor wireless is closely related to the environment, as its signal is easily blocked, reflected, and scattered by walls and other indoor objects \cite{BWP}. Therefore, the fundamental challenges of FAS are on how to accurately model the indoor environment-aware channel and then configure FAS accordingly, which nevertheless remains unsolved. 

Research on indoor wireless communications has been a thriving research area. For example, in \cite{Martin}, under statistical channel models, the authors illustrated that indoor wireless communication performance is significantly affected by wall blockages and derived analytical expressions for the average attenuation and for the signal-interference-ratio (SIR) performance. The authors of \cite{BWP-1} proposed closed-form expressions to evaluate building wireless performance through two figures of merit, namely interference and power gains, analyzing how room dimensions impact wireless performance at different frequencies to provide guidance for wireless-friendly building design. Aside from statistical channel modeling, a site-specific and accurate channel modeling approach, namely ray-tracing (RT), has been widely investigated for indoor wireless communications \cite{Fuschini, Yixin-1, Yixin-2, Chong, NYURay, hoydis2022sionna}.

The authors of \cite{Fuschini} highlights that RT approach can provide accurate channel models for indoor wireless communications, and can be used for MIMO performance assessment, beamforming optimization, and real-time channel prediction. Then in \cite{Yixin-1}, the authors examined how building materials affect indoor line-of-sight (LoS) MIMO wireless communications by developing a RT  model that incorporates a wall reflection. Moreover, the authors of \cite{Yixin-2} developed a RT model for square layouts under single-reflection condition, considering both LoS and non-LoS (NLoS) rays to evaluate how building materials affect MIMO communications. Furthermore, it was found in \cite{Chong} that LoS paths and wall-reflection paths dominate terahertz (THz) indoor channels through extensive measurements in a meeting room environment at $130$-$143$ GHz.

Unlike traditional Fresnel's reflection model, \cite{NYURay} proposed an angle-independent constant reflection loss model for RT, which allows for a closed-form linear least squares solution when calibrating the ray tracer to real-world measurements. But real-world data are usually expensive to obtain.  Sionna RT offers precise electromagnetic environment modeling with frequency-dependent material properties, and comprehensive propagation mechanisms including LoS paths, reflections, transmissions, and diffractions \cite{hoydis2022sionna}. Sionna's high-fidelity simulation capability comes at the cost of significant computational demands, often requiring GPU acceleration for efficient processing. By leveraging only a limited number of channel sounding measurements, a scatterer-based channel modeling framework was recently introduced in \cite{Sun2025}. In contrast to RT approach, it estimates large-scale fading by explicitly recovering the response coefficients of dominant scatterers in a specific region. However, to date, no analytical indoor site-specific model for FAS exists that strikes a tradeoff between high fidelity and computational complexity.

\begin{table*}[t]
	\centering
	\caption{\textsc{Definitions of Symbol Notation}}\label{tab:notation}
	\begin{tabular}{l|l|l|l|l|l}
		\hline
		\textbf{Symbol} & \textbf{Definition} & \textbf{Symbol} & \textbf{Definition} & \textbf{Symbol} & \textbf{Definition} \\ \hline
		$N$ & Number of transmit antennas & $K$ & Number of receivers & $\theta_j$ & Polar angle of transmit antenna $j$ \\ \hline	
		$M$ & Number of walls & $P_{\max}$ & Maximal transmit power  &  $\theta_l$ &  Left boundary of FAS\\ \hline
		$(x^c_1,y^c_1)$ & Coordinate of corner point $1$ &  $(x_0,y_0)$ & Coordinate of Tx  & $\theta_r$ &  Right boundary of FAS \\ \hline  
		$(x_1,y_1)$ & Coordinate of $\text{Rx}_1$ &  $\sigma^2$ & Variance of AWGN &  $\Delta$ & Minimal $\theta$ between adjacent FAS antennas \\ \hline  
		$\alpha$ & Angle of incidence & $S$ & Reflection point on the wall & $d_\text{LoS}$ & The length of LoS link\\ \hline  
	$d_\text{NLoS}$	& The length of NLoS link & $\bm{\omega}_1$ & The beamforming vector for $\text{Rx}_1$ & $\textbf{h}_1$ & The channel vector from $\text{Tx}$ to $\text{Rx}_1$ \\ \hline  
 $\pi$	&  MDP Policy & $\textbf{s}$ & MDP State & $\textbf{a}$ & MDP Action \\ \hline  
	$\textbf{o}$ & MDP observation &  	$r$ & MDP reward & $O$ & Trajectory length \\ \hline  	 
 $G$  & Group size & $T$ & Number of steps for PPO init. & $\epsilon_i$	& Relative permittivity of wall $i$ \\ \hline	 
	\end{tabular}
\vspace{-3mm}	
\end{table*}

The optimization of FAS configurations has been extensively investigated in the literature. If discrete FAS positions, i.e., antenna port, a joint antenna port, beamforming, and transmit power optimization design for point-to-point MIMO communication was first investigated in \cite{New}, where rank-revealing QR decomposition, singular value decomposition (SVD), and water-filling have been used for antenna port, beamforming, and transmit power optimization, respectively. Thereafter, the authors of \cite{Mei} considered the downlink multi-user MISO FAS with discrete FAS positions, where graph theory was applied to address the antenna port optimization problem.  Based on the derived outage probabilities, \cite{Yao} optimized the throughput of the reconfigurable intelligent surface (RIS)-assisted FAS system.

Breaking the discrete FAS positions assumption, \cite{Cheng,Qin} studied downlink multi-user MISO FAS with continuous FAS positions, where the FASs are equipped at the transmitter \cite{Cheng} and receiver \cite{Qin}, respectively. Fractional-programming, zero-forcing (ZF), and successive convex approximation (SCA) algorithms have been designed for the joint antenna position, beamforming, and transmit power problem. 
Moreover, \cite{HaoT} and \cite{Ye} tackled the maximization problem of the sensing signal-to-noise (SNR) with communication constraints for fluid antenna-assisted integrated sensing and communication (ISAC) systems   by SCA and majorization-maximization
(MM), respectively.  Among optimization methods, deep reinforcement learning (DRL) is particularly powerful for solving complex decision-making problems with high-dimensional state and action spaces. In \cite{Chao}, an advantage actor-critic (A2C) DRL algorithm was applied to jointly optimize antenna position, beamforming, and transmit power in FAS for ISAC. 

In this paper, we propose a layout-specific channel model, and a group relative policy optimization (GRPO) DRL solution for handling the coordinate optimization of antenna position, beamforming, and transmit power. A comparison of this work and related studies on FAS is given in Table \ref{tab:compare}. Our contributions are summarized as follows:
\begin{itemize}
\item \textit{Layout-Specific FAS Model}: We propose a layout-specific channel model for the indoor FAS. Specifically, for a layout, we introduce the indicator function to label NLoS and LoS rays, and design a low-complexity algorithm for function computation.  After that, we model the fluid antenna position using polar coordinates and analytically integrate the polar angles into the FAS channel model, leveraging the layout geometry. This method analytically provides a closed-form relationship between antenna movement and its impact on the channel model, enabling analytical analysis and concise optimization formulation for FAS. Moreover, a key advantage of our model is that it calculates wall reflections only once, significantly reducing computational complexity.  Compared to the start-of-the-art Sionna model, simulation results show that our model achieves an $83.3$\% computational time  reduction at the cost of approximately $3$ dB root-mean-square-error (RMSE), highlighting the merit for practical use.

\item \textit{Closed-Form Solution for Two-Ray FAS Model}: We simplify our proposed layout-specific model to the two-ray model. Although the antenna position optimization for rate maximization is a non-convex problem, we derive a closed-form solution by exploiting the insights in SNR function, namely the dominance and minimal length of the LoS path are important. Simulation results show that our closed-form solution delivers near-optimal performance across various problem settings, including rectangular layouts.

\item \textit{GRPO Solution for FAS}: For the joint optimization of antenna position, beamforming, and transmit power, we propose a GRPO solution that starts from a short-term trained proximal policy optimization (PPO) policy. Unlike PPO requires both A2C networks, GRPO optimizes memory utilization by leveraging group relative advantages, eliminating the need for the critic network. Extensive simulations show that our GRPO solution achieves a higher sum-rate than PPO, advantage A2C, and fixed antenna position with weighted minimum mean-square-error (WMMSE). Our GRPO solution requires only $50.8$\% of PPO's model size as well as floating-point operations per second (FLOPs).  
\end{itemize}
 
\textit{Organization}: The system model and problem formulation are presented in Section \ref{sec:model}. Then Section \ref{sec:RT} introduces the layout-specific RT model for FAS. A closed-form solution under a simplified single-wall reflection model is given in Section \ref{sec:closed-form}. GRPO solution for the joint optimization of antenna position, beamforming, and power allocation is proposed in Section \ref{sec:GRPO}. Comprehensive simulation results are provided in Section \ref{sec:result}. Conclusions are drawn in Section \ref{sec:conclude}.

\textit{Notation}:  $(\cdot)^*$ denotes conjugate operation. $(\cdot)^H$ denotes conjugate transpose operation. $\mathbb{E}\{\cdot\}$ denotes the long-term expectation operator. We write $f(x) = \mathcal{O}(g(x))$ as $x \to \infty$ if there exist constants $C > 0$ and $x' \in \mathbb{R}$ such that $\lvert f(x) \rvert \le C \lvert g(x) \rvert$ for all $x \ge x'$. Also, $[A]$ is a set containing elements from $1$ to $A$, and $\|\cdot\|_2$ is Euclidean norm. {The notation $\lfloor x \rfloor$ represents the floor of $x$, $\lceil x \rceil$ represents the ceiling of $x$, and $\lfloor x \rceil$ represents rounding $x$ to the nearest integer.}
Some notation definitions are given in Table \ref{tab:notation}. 

\section{System Model and Problem Formulation}\label{sec:model}
We consider a FAS deployed in a room, where all the walls are assumed to be either perpendicular or parallel. We also consider there are $K$ receivers, denoted by Rx, and a single transmitter, denoted by Tx, within this layout, where each Rx has a single antenna, while Tx has $N$ fluid antennas and $K \le N$. $\text{Tx}$ aims to communicate with all $\text{Rxs}$ simultaneously, thus creating undesired interference to each $\text{Rx}$. The additive white Gaussian noise (AWGN) is expressed as $\mathcal{CN}(0,\sigma^2)$.  

\begin{figure}[t]
	\centering
	\includegraphics[width=3in]{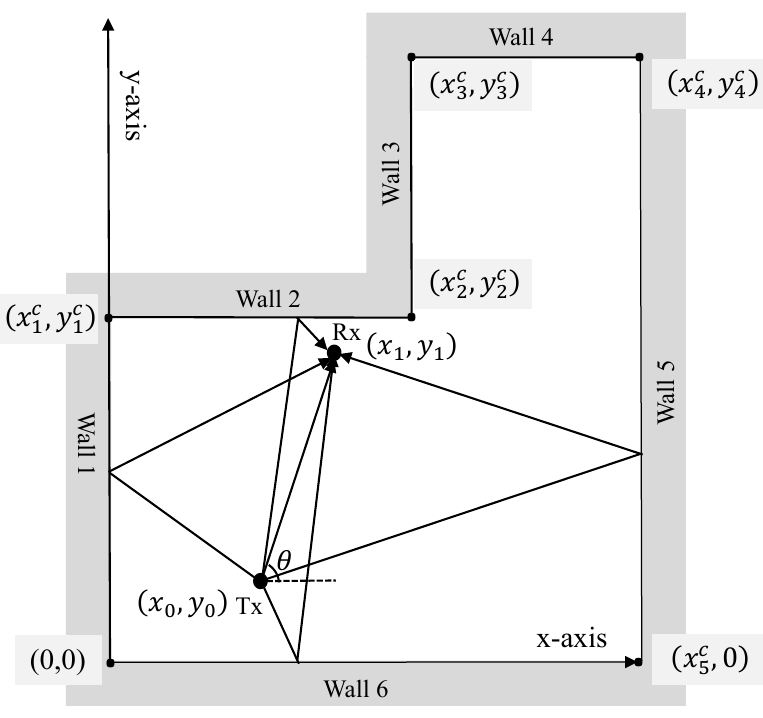}
	\caption{A room layout for exemplified illustration.} \label{F3}
\end{figure}

Specifically, there exists a room layout with $M$ walls, for radio propagation, there exist multiple rays from $\textsc{Tx}$ to $\text{Rx}$. An example of room layout is as shown in Fig. \ref{F3}. A square coordinate system with $x$-axis and $y$-axis is introduced. The locations of $\text{Rx}_k,$ and $\textsc{Tx}$ are denoted by $(x_k,y_k),$ and $(x_0,y_0)$, respectively, represented as a set $\mathcal{X} \triangleq \{x_0,y_0,x_1,y_1,\cdots,x_K,y_K\}$. {We assume a rectilinear layout, where all walls are either perpendicular or parallel to the $x$-axis.  This configuration is widely recognized as a preliminary and predominant layout in the design of rectangular architectural layouts \cite{sharafi2015conceptual}. For convenience, we describe this layout by specifying the corner points of each wall.} The corner points in clockwise order are denoted by $(0,0),(x_1^c,y_1^c),\ldots,(x_{M-1}^c,0)$, respectively, and represented as a set $\mathcal{L} \triangleq \{0,0,x^c_1,y^c_1,\cdots,x^c_{M-1},0\}$. Given room layout $\mathcal{L}$ and $\text{Tx}$-$\text{Rx}$ locations $\mathcal{X}$, the channel from $\text{Tx}$ to $\text{Rx}$ $k$ is parameterized by polar angle $\theta_{k,j}$ of Tx-antenna $j$'s position, denoted by $\mathbf{h}_k(\bm{\theta}) = [h_{k,1}(\theta_{1}),\cdots,h_{k,N}(\theta_{N})]^T$, where $\bm{\theta} = (\theta_1,\cdots,\theta_N)$ with $(\theta_{k,1},\cdots,\theta_{k,N}) = \bm{\theta} + \bm{\delta}_k$ and $\bm{\delta}_1 = \mathbf{0}$. $\text{Tx}$ transmits one data stream to each $\text{Rx}$ with beamforming vector $\bm{\omega}_k$. The signal-to-interference-noise ration (SINR) at $\text{Rx}$ $k$ is expressed as $\frac{|\textbf{h}_k^H(\bm{\theta}) \bm{\omega}_k|^2}{\sum_{k' \ne k} |\textbf{h}_{k'}^H(\bm{\theta})\bm{\omega}_k|^2 + \sigma^2}$, and the data rate at $\text{Rx}$ $k$ is expressed as $\log_2(1 + \frac{|\textbf{h}_k^H(\bm{\theta}) \bm{\omega}_k|^2}{\sum_{k' \ne k} |\textbf{h}_{k'}^H(\bm{\theta})\bm{\omega}_k|^2 + \sigma^2})$. Our objective is to maximize the sum-rate through joint optimization of the fluid antenna positions, power allocation, and beamforming vectors. This optimization problem is formulated as follows:
\begin{subequations}
	\begin{eqnarray}
		\!\!\!\!\!\!\!\!\!\!	(\text{P1}) \,\,\, \max_{\bm{\theta}, \{\bm{\omega}_k\}} && \!\!\!\!\!\! \sum_{k=1}^K \log_2\left(1 + \frac{|\textbf{h}_k^H(\bm{\theta}) \bm{\omega}_k|^2}{\sum_{k' \ne k} |\textbf{h}_{k'}^H(\bm{\theta})\bm{\omega}_k|^2 + \sigma^2}\right) \nonumber \\
		\!\!\!\!\!\!\!\!\!\!	\text{s.t.} &&  \!\!\!\!\!\! \sum_{k=1}^K \| \bm{\omega}_k \|_2^2 \le P_{\max}, \label{24a} \\
		\!\!\!\!\!\!\!\!\!\!	&&  \!\!\!\!\!\! \theta_l \le \theta_1 \le \theta_2 \le \cdots \le \theta_K \le \theta_r, \label{24b} \\
		\!\!\!\!\!\!\!\!\!\!	&&  \!\!\!\!\!\! \Delta \le |\theta_{k'} - \theta_{k''}|, \quad \forall k', k'' \in [K], \label{24c}
	\end{eqnarray}
\end{subequations}
where constraint \eqref{24a} ensures the total transmission power remains within the allowable limit $P_{\max}$, constraint \eqref{24b} guarantees that the optimized FAS antenna positions are sequentially ordered, and constraint \eqref{24c} enforces a minimum separation $\Delta$ between adjacent FAS antennas to avoid mutual coupling.

Prior to solving Problem (P1), we shall establish the model of $\mathbf{h}_k(\bm{\theta})$. This model should satisfy two key requirements:  1) the ability to capture the multi-rays propagating in a room layout; 2) the channel should depend uniquely on the parameters $\theta_{j}$. Next section, we propose a model that fulfills the aforementioned requirements.

\section{Proposed Layout-Specific RT Model for FAS}\label{sec:RT}

To model the layout-specific RT channel, we consider the ray reflection on each wall only once. After determining whether there is a ray reflection on each wall, we analytically express the length of NLoS and LoS rays. Given a room layout $\mathcal{L}$ and and \text{Tx-Rxs} locations $\mathcal{X}$, we model this layout specific RT channel as $\mathbf{h}_k(\mathcal{L},\mathcal{X}) = (h_{k,1}(\mathcal{L},\mathcal{X}),\cdots,h_{k,N}(\mathcal{L},\mathcal{X}))$, where $h_{k,j}(\mathcal{L},\mathcal{X})$ denotes the per-antenna channel from Tx-antenna $j$ to $\text{Rx}_k$, modeled in \eqref{Th2}, shown on the top of next page, where $\lambda$ denotes the wavelength; $\epsilon_i$ denotes the relative permittivity of wall $i$; $d_{\text{LoS},j}$ denotes the length of LoS ray transmitted from Tx-antenna $j$; $d_{\text{NLoS-}i,k,j}$ denotes the length of NLoS ray transmitted from Tx-antenna $j$, reflected by wall $i$, and arriving at $\text{Rx}_k$; $\alpha_{i,k,j}$ denotes the angle of incidence for NLoS Ray transmitted from Tx-antenna $j$, reflected by wall $i$, and arriving at $\text{Rx}_k$; $\Gamma_{i,k,j}$ denotes the reflection loss for NLoS Ray transmitted from Tx-antenna $j$, reflected by wall $i$, and arriving at $\text{Rx}_k$. The wavelength $\lambda$ is calculated by $3 \times 10^8 / f$, where $f$ denotes the frequency. Specifically, we introduce indicator functions in \eqref{Th2} to indicate the absence or presence of rays.

\begin{algorithm}[!t]
	\caption{Indicator Functions Computation}
	\label{algorithm:ray_detection}
	\KwIn{Tx-Rx set $\mathcal{X}$, Wall set $\mathcal{L}$}
	
	Set $1$ to $\{\mathds{1}_{\text{Wall-}i,k,j}(\mathcal{L}, \mathcal{X})\}$, and  to $\mathds{1}_{\text{LoS},j}(\mathcal{L}, \mathcal{X})$.\\
	
	\textbf{Step 1: Compute Reflection Points $\{S_i\}$ in $\mathcal{L}$} \\
	\For{each Wall $i$ in $\mathcal{L}$}{
		\If{wall $i$ is perpendicular to $y$-axis}{
			Compute \eqref{17} for $u$. $S_i = (x_0+u,y_i^c)$.
		}
		\If{wall $i$ is perpendicular to $x$-axis}{
			Compute \eqref{17} for $u$. $S_i = (x_i^c,y_0+u)$.
		}	
	}  
	\textbf{Step 2: Identify Blockage in $\mathcal{L}$ for NLoS Rays}\\
	\For{each Wall $i$ in $\mathcal{L}$}{
		Line equation $\ell_i^{[1]}$ generation by $(x_0,y_0)$ and $S_i$. \\
		Line equation $\ell_i^{[2]}$ generation by $(x_1,y_1)$ and $S_i$. 
		\For{each wall $i'$ in $\mathcal{L}$}{
			\If{The linear system $\ell_i^{[1]}$, $x = x_{i'}^c-x_{{i'}-1}^c$, and $y=y_{i'}^c-y_{{i'}-1}^c$ has a non-zero solution}
			{$\mathds{1}_{\text{Wall-}i,k,j}(\mathcal{L},\mathcal{X}) = 0$.}
			\If{The linear system $\ell_i^{[2]}$, $x = x_{i'}^c-x_{{i'}-1}^c$, and $y=y_{i'}^c-y_{{i'}-1}^c$ has a non-zero solution}
			{$\mathds{1}_{\text{Wall-}i,k,j}(\mathcal{L},\mathcal{X}) = 0$.}
	}}
	\textbf{Step 3: Identify Blockage in LoS Ray} \\
	Line equation $\ell_i^{[3]}$ generation by $(x_0,y_0)$ and $(x_1,y_1)$.
	\For{each wall $i'$ in $\mathcal{L}$}{
		\If{The linear system $\ell_i^{[3]}$, $x = x_{i'}^c-x_{{i'}-1}^c$, and $y=y_{i'}^c-y_{{i'}-1}^c$ has a non-zero solution}
		{$\mathds{1}_{\text{LoS},j}(\mathcal{L},\mathcal{X}) = 0$.}
	}

	\KwOut{$\{\mathds{1}_{\text{Wall-}i,k,j}(\mathcal{L}, \mathcal{X})\}$, $\mathds{1}_{\text{LoS},j}(\mathcal{L}, \mathcal{X})$}
\end{algorithm}

The indicator function is represented by $\mathds{1}_{\text{Wall-}i,k,j}(\mathcal{L},\mathcal{X})$ for $i \in [M],\,k \in [K],\,j \in [N]$, where $1$ indicates the existence of an NLoS ray of wall $i$, and $0$ indicates its absence.  Similarly, we define $\mathds{1}_{\text{LoS},j}(\mathcal{L},\mathcal{X})$.
Given a room layout, all indicator functions can be determined using Algorithm~1.
The computational complexity of Algorithm 1 can be assessed across its three steps, with $M$ walls in the set $\mathcal{L}$ and a fixed Tx-Rx pair in $\mathcal{X}$. Step 1 checks each wall in $\mathcal{O}(M)$. Step 2 computes reflection points for up to $M$ walls, also taking $\mathcal{O}(M)$. Step 3, the most demanding, involves checking each of the filtered walls up to $M$ against all $M$ walls for blockages, resulting in $\mathcal{O}(M^2)$. Step 4 checks the LoS against all $M$ walls in $O(M)$. Since Step 3 dominates, the overall computational complexity of Algorithm 1 is $\mathcal{O}(M^2)$.

\begin{figure*}
	\begin{eqnarray}
	&& h_{k,j}(\mathcal{L},\mathcal{X}) = \frac{\sqrt{G_tG_r}\lambda}{4\pi }  \left(\underbrace{\sum_{i=1}^M \mathds{1}_{\text{Wall-}i,k,j}(\mathcal{L},\mathcal{X})\frac{\Gamma_{i,k,j} \exp(-j2\pi \frac{d_{\text{NLoS-}i,k,j}}{\lambda})}{d_{\text{NLoS-}i,k,j}}}_{\text{NLoS Ray}}  
		+ \underbrace{\mathds{1}_{\text{LoS},j}(\mathcal{L},\mathcal{X})\frac{\exp(-j2\pi\frac{ d_{\text{LoS},j}}{\lambda})}{d_{\text{LoS},j}}}_{\text{LoS Ray}}\right), \label{Th2} \\ 
	&& \text{where}	 \nonumber \\
	&& 	 	\Gamma_{i,k,j} = 
	 	\begin{cases}
		 		\dfrac{- \epsilon_i \sin \alpha_{i,k,j} + \sqrt{\epsilon_i - \cos^2 \alpha_{i,k,j}}}{\epsilon_{i} \sin \alpha_{i,k,j} + \sqrt{\epsilon_i - \cos^2 \alpha_{i,k,j}}}, & \text{{T}ransverse {M}agnetic (TM) Mode},	\\
		 		\dfrac{\sin \alpha_{i,k,j}-\sqrt{\epsilon_i - \cos^2 \alpha_{i,k,j}}}{\sin \alpha_{i,k,j} + \sqrt{\epsilon_i - \cos^2 \alpha_{i,k,j}}}, & \text{{T}ransverse {E}lectric (TE) Mode}.
	 	\end{cases} \label{GGamma}	
	\end{eqnarray}
	\hrule
\end{figure*}

Once the indicator functions are decided, we proceed to parameterize the length of NLoS ray $d_{\text{NLoS-}i,k,j}$, and the length of LoS ray $d_{\text{LoS},j}$, and Fresnel's reflection coefficient $\Gamma_{i,k,j}$ using Tx-antenna $j$'s polar angle $\theta_{k,j}$. This is through introducing {angle-dependent} distance functions $D^{[1]}_{i,k}(\theta_{k,j}),\,D^{[2]}_{i,k}(\theta_{k,j})$ and $D^{[3]}_{i,k}(\theta_{k,j})$, where $D^{[1]}_{i,k}(\theta_{k,j})$ and $D^{[2]}_{i,k}(\theta_{k,j})$ denote the perpendicular {Euclidean} distances from $\text{Tx}$ and $\text{Rx}_k$ to wall $i$, respectively, and $D^{[3]}_{i,k}(\theta_{k,j})$ denotes the horizontal Euclidean distance between $\text{Tx}$ and $\text{Rx}_k$. 

\begin{figure}[t]
	\centering
	\includegraphics[width=2.75in]{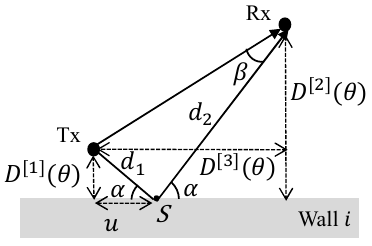}
	\caption{The two-ray model under reflection of wall $i$.} \label{F4}
\end{figure}

We now show the parameterization of the terms $d_{\text{NLoS-}i,k,j}$, $d_{\text{LoS},j}$, and $\Gamma_{i,k,j}$ with respect to the angle $\theta_j$. For notational simplicity, the subscripts are omitted in the subsequent analysis.
As revealed by the geometry in Fig.~\ref{F4}, the following equation holds,
\begin{subequations}
	\begin{eqnarray}
		&&	\tan \alpha = \frac{D^{[1]}(\theta)}{u}, \label{Q1} \\
		&&	\tan \alpha = \frac{D^{[2]}(\theta)}{D^{[3]}(\theta) - u}, \label{Q2} 
	\end{eqnarray}
\end{subequations}
where $u$ can be solved from \eqref{Q1}  and \eqref{Q2} as
\begin{equation}
	u = \frac{D^{[1]}(\theta)D^{[3]}(\theta)}{D^{[1]}(\theta) + D^{[2]}(\theta)}. \label{17}
\end{equation}
Next, we eliminate $u$ in \eqref{Q1}-\eqref{Q2} and obtain, 
\begin{equation}
	\qquad\,\,\,\,	\alpha = \arctan \frac{D^{[1]}(\theta) + D^{[2]}(\theta)}{D^{[3]}(\theta)}. \label{Q3} 
\end{equation}
The length of NLoS ray obeys the following relationship,
\begin{equation}
	(d_1 + d_2)\sin \alpha = D^{[1]}(\theta)+D^{[2]}(\theta), \label{QQQQ}
\end{equation}
Since $\sin \alpha = \frac{\tan \alpha}{ \sqrt{1 + \tan^2 \alpha}}$, \eqref{Q3}, and \eqref{QQQQ}, we arrive at 
\begin{equation}
	d_\text{NLoS} = d_1 + d_2 =  \sqrt{ {D^{[3]}}^2(\theta) + (D^{[1]}(\theta) + D^{[2]}(\theta))^2},  \label{Q4}
\end{equation}
Furthermore, we express Fresnel's reflection coefficient $\Gamma$ with $D^{[1]}(\theta),\,D^{[2]}(\theta)$ and $D^{[3]}(\theta)$.

For an arbitrary wall $i$ in the layout $\mathcal{L}$, we can obtain its $\alpha$ and $d_{\text{NLoS}}$ by \eqref{Q3} and \eqref{Q4}, respectively.
Finally, it can be seen that
\begin{eqnarray}
	d_{\text{LoS}} = \begin{cases}
		\frac{y_1 - y_0}{\sin \theta}, & \theta \in (\arctan \frac{y_1 - y_0}{x_1}, \frac{\pi}{2}), \\
		\frac{y_1 - y_0}{\sin (\pi - \theta)}= \frac{y_1 - y_0} {\sin \theta}, & \theta \in [\frac{\pi}{2}, \pi - \arctan \frac{y_1 - y_0}{x_1}).
	\end{cases} \label{22}
\end{eqnarray}

As a final step, we shall derive the angle-dependent distances $D^{[1]}(\theta)$, $D^{[2]}(\theta)$, and $D^{[3]}(\theta)$ for a given room layout. This derivation is achieved by expressing these distances in terms of a unified coordinate representation, where the distance variables are parameterized by the polar angle $\theta$. Please refer to Appendix A for an example. Through $D^{[1]}(\theta)$, $D^{[2]}(\theta)$, $D^{[3]}(\theta)$, and Algorithm 1, $h_{k,j}(\theta_j)$ can be obtained via \eqref{Th2} and \eqref{GGamma}, thereby yielding the desired expression for $\mathbf{h}_k(\bm{\theta})$.
\begin{figure}[t]
    \centering
    \includegraphics[width=0.5\linewidth]{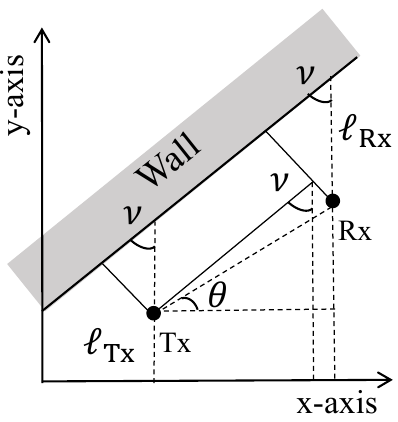}
    \caption{Illustration of the angle-dependent distance derivation for an oblique wall.}
    \label{MR-TWC}
\end{figure}

\subsection{Discussion on Spatial Correlation}
Moreover, we discuss the spatial correlation for the proposed channel model as follows. Unlike statistical FAS channel models in \cite{Mohamed-Slim, RostamiGhadi, RamirezEspinosa}, we model the spatial correlation in deterministic propagation environments by the RT method. Therefore, the spatial correlation is inherently characterized by the ray propagation \cite{Fuschini}. According to \cite{Fuschini}, the correlation between the channel from transmit antenna $j_1$ to $\text{Rx}_k$, i.e., $h_{k,j_1}(\theta_{j_1})$ and the channel from transmit antenna $j_2$ to $\text{Rx}_k$, i.e., $h_{k,j_2}(\theta_{j_2})$ can be calculated by
\begin{equation}
    \rho_{1,2} = \frac{h_{k,j_1}(\theta_{j_1})h_{k,j_2}^*(\theta_{j_2})}{|h_{k,j_1}(\theta_{j_1})|^2|h_{k,j_2}(\theta_{j_2}))|^2},
\end{equation}
where $0 \le |\rho_{1,2}| \le 1$, $|\rho_{1,2}| = 1$ represents a linear correlation and $|\rho_{1,2}| = 0$ represents no correlation.

\subsection{{Discussion on Arbitrary Geometry Layout}}

In the presence of oblique walls, shown in Fig. \ref{MR-TWC}, the changes lie in deriving the angle-dependent distances $D^{[1]}(\theta)$, $D^{[2]}(\theta)$, and $D^{[3]}(\theta)$ as well as indicator function computation algorithm, such that $\mathbf{h}_k(\bm{\theta})$ can be obtained via \eqref{Th2} and \eqref{GGamma}. The complexity arises from the wall’s slope, which introduces nontrivial geometric and dependencies into both the distance functions and the indicator function computation. Specifically, for the angle-dependent distance derivation, we shall first derive the intersection points of the upward prolongation lines $\ell_\text{Tx}$, $\ell_\text{Rx}$, and the wall. After that, we can have the distance between the intersection point and Tx or Rx. Finally, since slope angle $\nu$ is known, $D^{[1]}(\theta)$ and $D^{[2]}(\theta)$ can be derived. Since the distance between Tx and Rx can be obtained by $d_\text{LoS}$, $D^{[3]}(\theta)$ can be derived as $d_\text{LoS}\cos(\frac{\pi}{2}-\theta-\nu)$.
For the indicator function computation, the wall’s slope must be considered when computing reflection points. To properly detect blockages, the oblique wall equation replaces the corresponding perpendicular or parallel to the $x$-axis equation.

In the presence of curved walls, this scenario becomes significantly more complex. The RT algorithm proposed in Sionna \cite{hoydis2022sionna} should be employed, where both reflection point computation and blockage identification are handled algorithmically rather than through semi-closed-form expressions.
 
\begin{figure}[t]
	\begin{subfigure}{0.49\linewidth}
		\centering
		\includegraphics[width=0.8\linewidth]{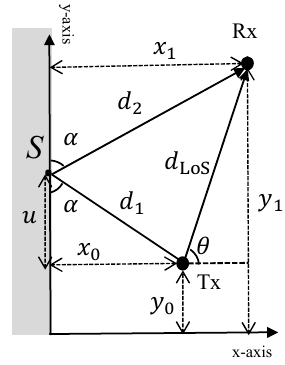}
		\caption{When $\arctan \frac{y_1 - y_0}{x_1}  \le \theta < \frac{\pi}{2}$}
	\end{subfigure} 
	\hfill
	\begin{subfigure}{0.49\linewidth}
		\centering
		\includegraphics[width=\linewidth]{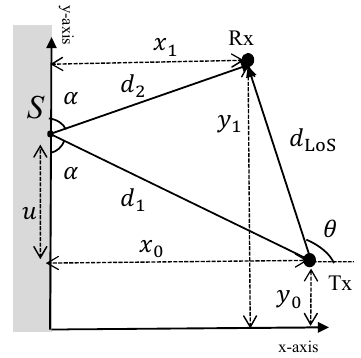}
		\caption{When $\frac{\pi}{2} \le \theta < \pi$}	 
	\end{subfigure}
	\caption{Two-ray model (single-wall reflection only), where the reflection point on the wall is denoted by $S$.} \label{F1}
\end{figure}

\begin{figure*}[t]
	\centering
	\begin{subfigure}{0.24\linewidth}
		\centering
		\includegraphics[width=0.96\linewidth]{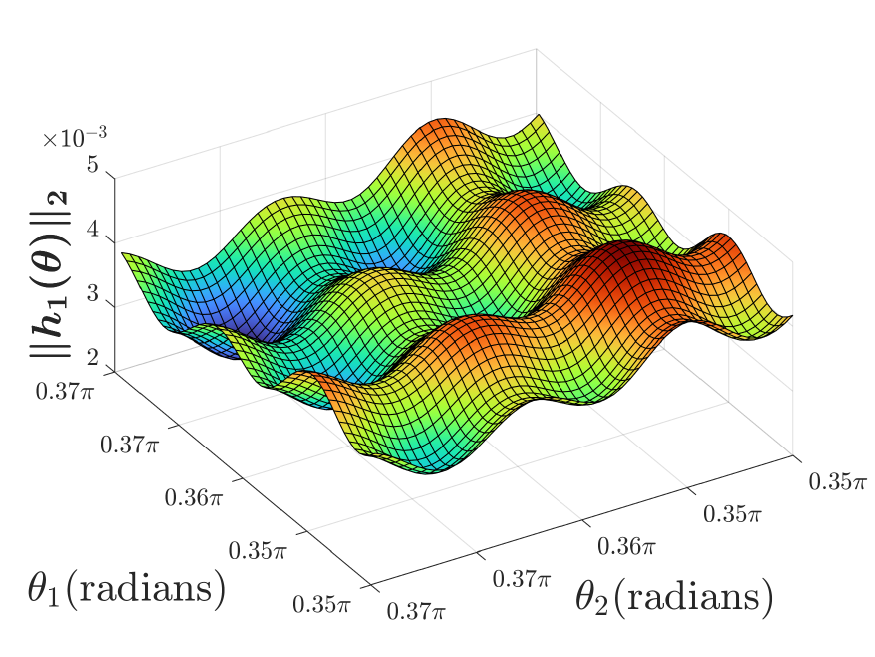}
		\caption{$\text{Rx}_1,\,x_1=y_1=2.5$m. { \\$f = 5\,\text{GHz}$}}	 
	\end{subfigure}
	\centering
	\begin{subfigure}{0.24\linewidth}
		\centering
		\includegraphics[width=0.96\linewidth]{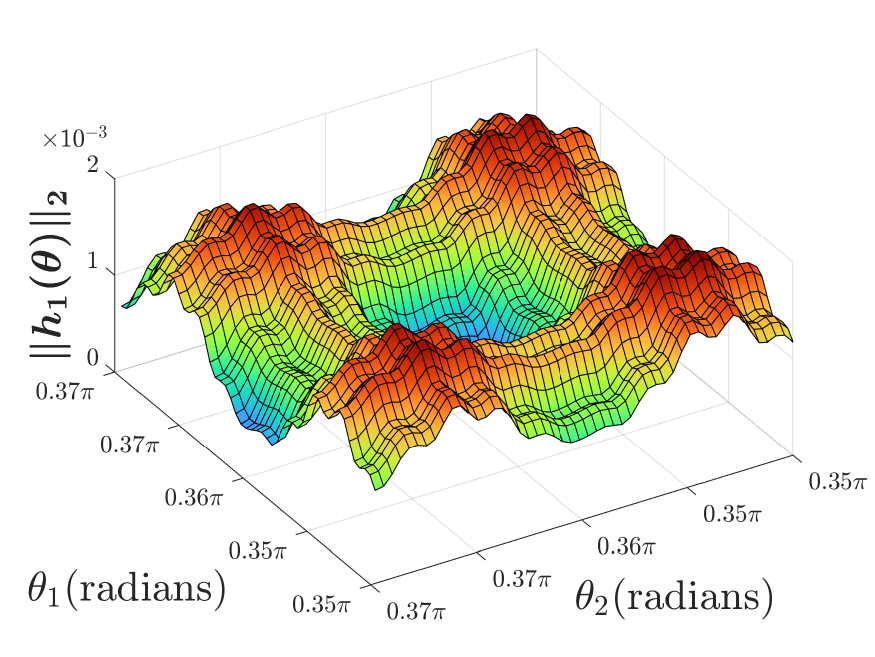}
		\caption{$\text{Rx}_2,\,x_2=8.5$m, $y_2=6$m. { $f = 5\,\text{GHz}$}}		 
	\end{subfigure}
	\centering
	\begin{subfigure}{0.24\linewidth}
		\centering
		\includegraphics[width=0.96\linewidth]{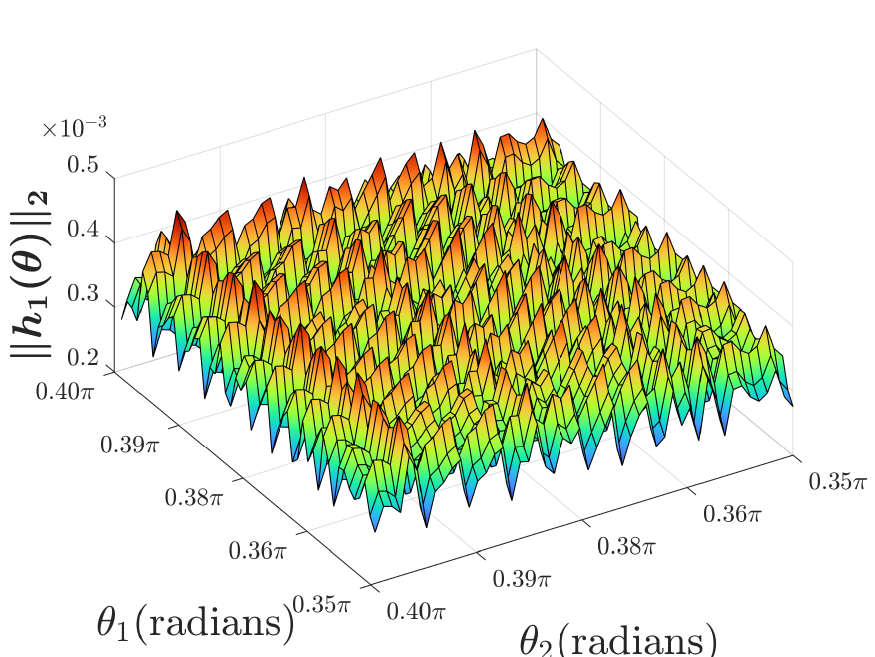}
		\caption{{$\text{Rx}_1,\,x_1=y_1=2.5$m. \\ $f = 60\,\text{GHz}$}}	 
	\end{subfigure}
	\centering
	\begin{subfigure}{0.24\linewidth}
		\centering
		\includegraphics[width=0.96\linewidth]{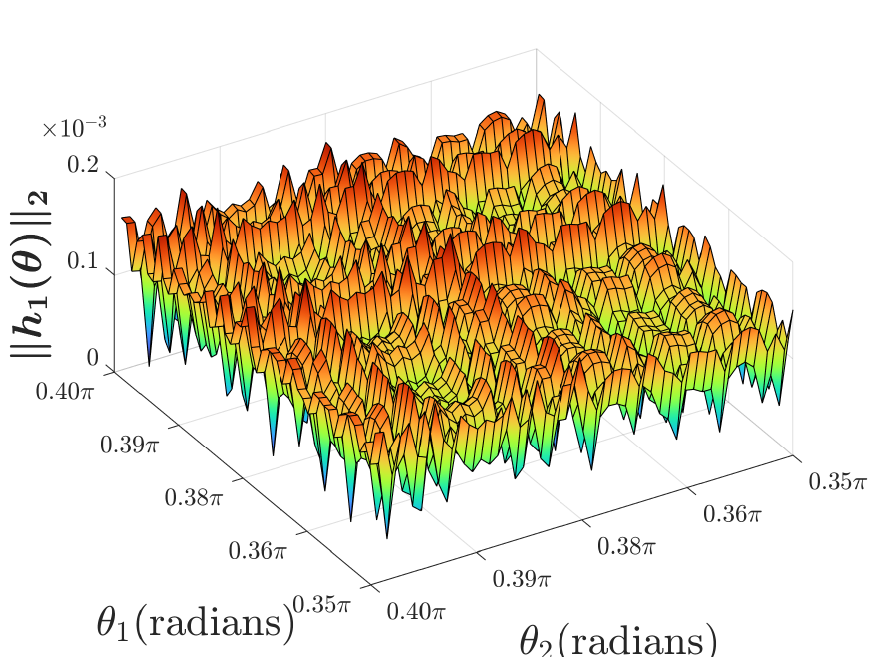}
		\caption{{$\text{Rx}_2,\,x_2=8.5$m, $y_2=6$m.  $f = 60\,\text{GHz}$}}		 
	\end{subfigure}

	\centering
	\begin{subfigure}{0.24\linewidth}
		\centering
		\includegraphics[width=0.96\linewidth]{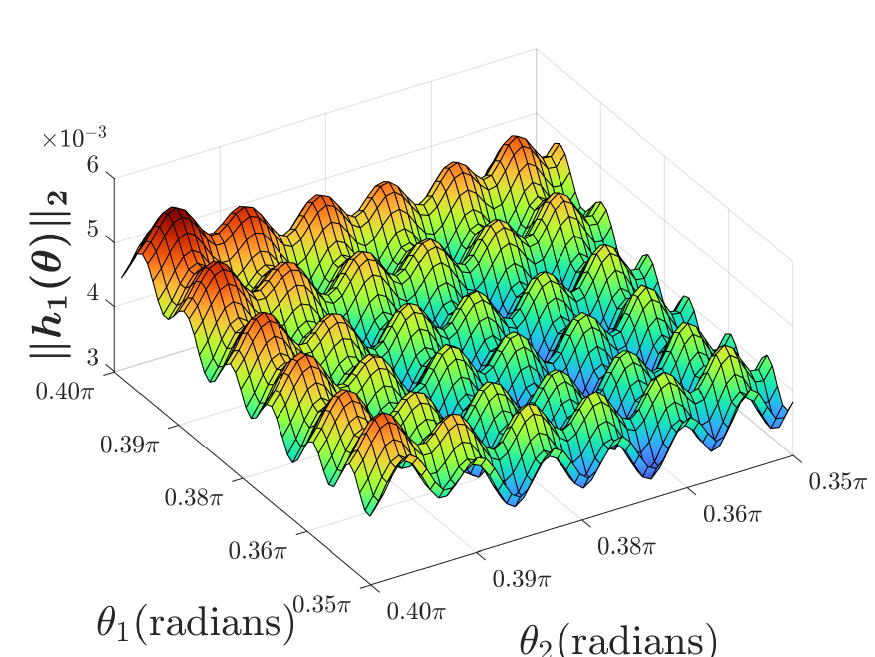}
		\caption{{$\text{Rx}_1,\,x_1=y_1=2.5$m. \\$f = 5\,\text{GHz}$}}	 
	\end{subfigure}
	\centering
	\begin{subfigure}{0.24\linewidth}
		\centering
		\includegraphics[width=0.96\linewidth]{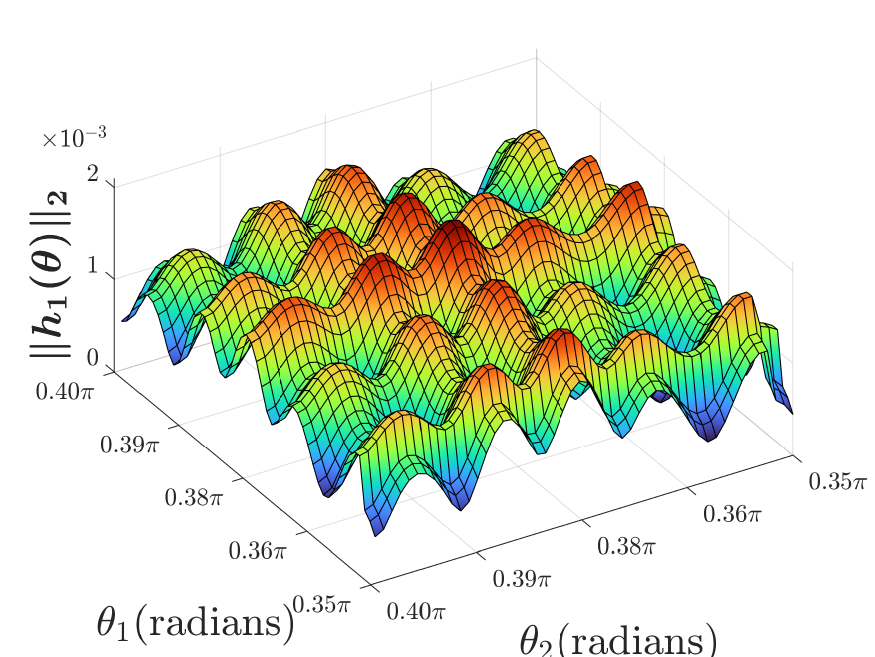}
		\caption{{$\text{Rx}_2,\,x_2=8.5$m, $y_2=6$m. $f = 5\,\text{GHz}$}}		 
	\end{subfigure}
	\centering
	\begin{subfigure}{0.24\linewidth}
		\centering
		\includegraphics[width=0.96\linewidth]{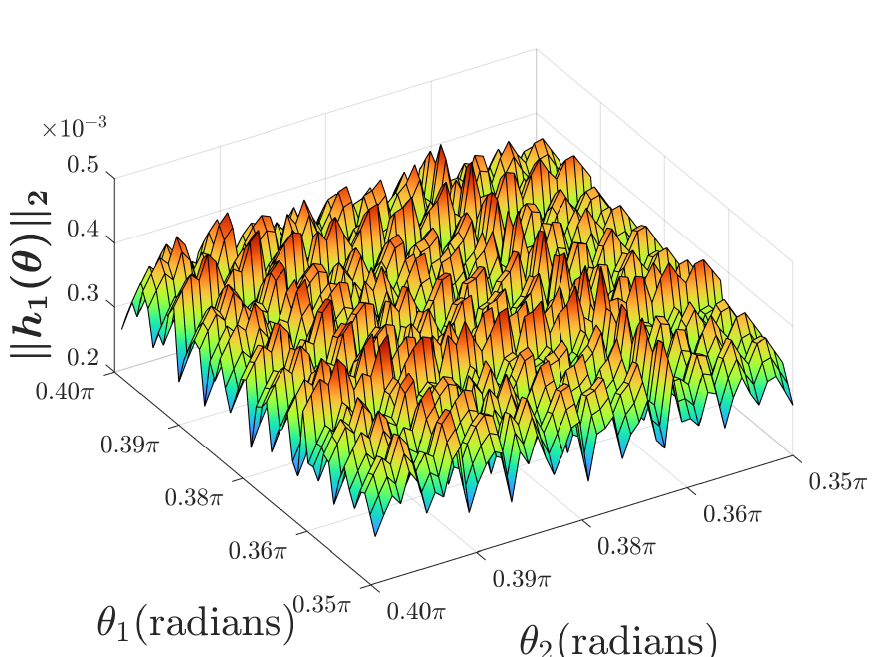}
		\caption{{$\text{Rx}_1,\,x_1=y_1=2.5$m. \\$f = 60\,\text{GHz}$}}	 
	\end{subfigure}
	\centering
	\begin{subfigure}{0.24\linewidth}
		\centering
		\includegraphics[width=0.96\linewidth]{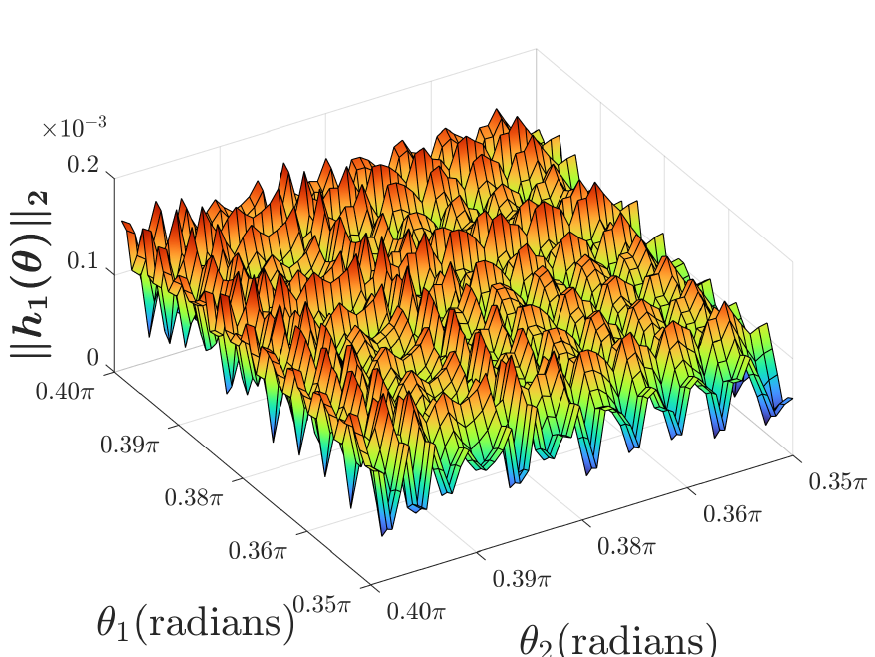}
		\caption{{$\text{Rx}_2,\,x_2=8.5$m, $y_2=6$m. $f = 60\,\text{GHz}$}}		 
	\end{subfigure}
	\caption{ Landscape of $\|\textbf{h}_k({\bm{\theta}})\|_2, k=1,2,$ in TE mode, $G_t=G_r=1, \sigma^2 = -90\,\text{dBm},  y_0=1$m, $ f = 5\,\text{GHz},\epsilon =  5.24$. Subfigures (a)--(d): Rectangular layout with corner points $\{(0,0), (0,10), (10,10), (10,0)\}$ (Unit: m). Subfigures (e)--(h): Irregular layout following Fig.~1, 		with corner points $\{(0,0), (0,5), (5,8), (8,8), (8,10), (10,10), (10,0)\}$ (Unit: m).}
	\label{5}
\end{figure*}

\section{Closed-From Solution for \\ Two-Ray Model}\label{sec:closed-form}
 
It can be seen that Problem (P1) poses a significant challenge due to its non-convex nature, stemming from the channel model in the objective function. {To show this issue edgewise, we show the landscape of $\|\textbf{h}_1(\bm{\theta})\|_2$ and $\|\textbf{h}_2(\bm{\theta})\|_2$ in Fig. \ref{5} for the rectangular layout and Fig. \ref{F3} layout with $N=2$ and $K=2$.} As illustrated in Fig. \ref{5}, the landscape reveals a multitude of local optima, which complicates the application of conventional numerical optimization techniques and hinders their ability to achieve a high-performing solution. Therefore, in this section, we first simplify the channel model from the layout-specific downlink multi-user MISO to the single-wall reflection with single-antenna point-to-point transmission, i.e., two-ray. By this way, we can have a reduced form of  Problem (P1) and next obtain a closed-form solution by leveraging some key insights.

\begin{figure*}
		\begin{eqnarray}
	  \textsc{SNR}_{2}(\theta) =  \frac{G_tG_r\lambda^2}{\sigma^2\left(4\pi\right)^2}     \left( \underbrace{\frac{\Gamma^2}{(y_1 - y_0)^2 + (2x_1 - \frac{y_1 - y_0}{\tan \theta})^2}}_{\text{NLoS Term}} + \underbrace{\frac{\sin^2 \theta}{(y_1 - y_0)^2}}_{\text{LoS Term}}  
	  +   \underbrace{\frac{2 \Gamma \cos \left( 2\pi \frac{\sqrt{(y_1 - y_0)^2 + (2x_1 - \frac{y_1 - y_0}{\tan \theta})^2} - \frac{y_1 - y_0}{ \sin \theta}}{\lambda}\right)}{ \frac{y_1 - y_0}{ \sin \theta} \sqrt{(y_1 - y_0)^2 + (2x_1 - \frac{y_1 - y_0}{\tan \theta})^2}}}_{\text{NLoS-LoS Interaction Term}}  
		\right),  
		\label{RG10}
	\end{eqnarray}
		\hrule 
\end{figure*}

If we assume that there are only one Rx with a single antenna at Tx. Furthermore, we assume that there exists only a wall rather than a layout for simplicity. As shown in Fig. \ref{F1}, we have one Rx, whose location is denoted by $(x_1,y_1)$. The wall is perpendicular to $x$-axis. The channel is given by
\begin{eqnarray}
	&& \!\!\!\!\!\!\!\!\!\!\!\!\!\!\!\! h_2(\mathcal{L},\mathcal{X}) = \nonumber \\
	&& \!\!\!\!\!\!\!\!\!\!\!\!\!\!\!\! \frac{\sqrt{G_tG_r}\lambda}{4\pi }  \left(\underbrace{\frac{\Gamma \exp(-j2\pi \frac{d_{\text{NLoS}}}{\lambda})}{d_{\text{NLoS}}}}_{\text{NLoS Ray}}  
	+ \underbrace{\frac{\exp(-j2\pi\frac{ d_{\text{LoS}}}{\lambda})}{d_{\text{LoS}}}}_{\text{LoS Ray}}\right),
\end{eqnarray}
where $\Gamma$ is from \eqref{GGamma}.
The key to express the channel with $\theta$ is writing down $D^{[1]}(\theta)$,  $D^{[2]}(\theta)$, and $D^{[3]}(\theta)$. After that, we are able to arrive at $h_2(\theta)$ using {\eqref{Q3}, \eqref{Q4},  and \eqref{22}}. Finally, $\text{SNR}_2(\theta) = |h_2(\theta)|^2/\sigma^2$. The expression of $\text{SNR}_2(\theta)$ is given by below Theorem.

\begin{theorem}
Given the simplified single-wall reflection model with point-to-point transmission, SNR function is given in \eqref{RG10}, shown on the top of next page, where $\theta \in (\arctan \frac{y_1 - y_0}{x_1}, \pi)$ and $\Gamma$ is from substituting $\alpha  = \arctan ( \frac{2x_1}{y_1 - y_0} - \frac{1}{\tan \theta})$ into \eqref{GGamma}. 
\end{theorem}

	\begin{IEEEproof}
		Please refer to Appendix B.
	\end{IEEEproof}
	It can be observed from \eqref{RG10} that there are three terms in $\textsc{SNR}_{2}(\theta)$: the first term is exclusively influenced by the NLoS ray, the second term is solely affected by the LoS ray, and the third term represents the interaction/interference between NLoS and LoS rays.  For this simplified problem, Problem (P1) reduces to, 
	\begin{subequations}
		\begin{eqnarray}
			(\text{P2}) \quad \max_{\theta} && \log_2\left(1 + \text{SNR}_{2}(\theta)\right)  \\
			\text{s.t.} && \theta_l \leq \theta \leq \theta_r, 	
		\end{eqnarray}
	\end{subequations}
	where \(\theta\) can be varied from the lower bound \(\theta_l\) to the upper bound \(\theta_r\). In an interference-free scenario, maximizing \(\log_2\left(1 + \text{SNR}_{2}(\theta)\right)\) is equivalent to maximizing \(\text{SNR}_{2}(\theta)\).

\begin{figure}[t]
		\centering
		\begin{subfigure}{0.475\linewidth}
			\centering
			\includegraphics[width=0.97\linewidth]{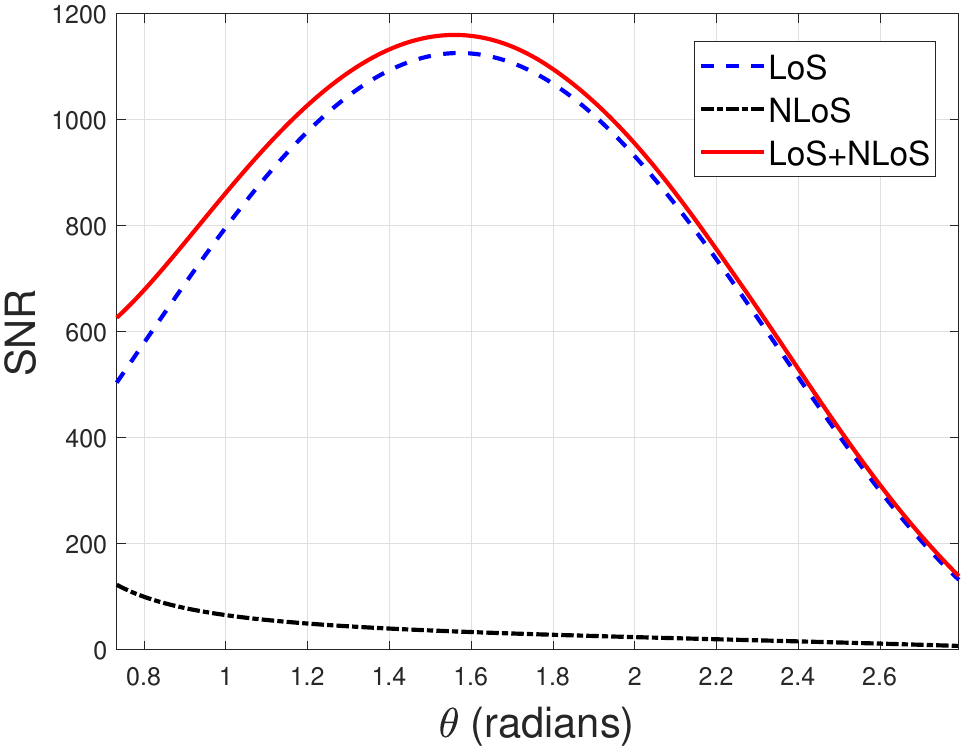}
			\caption{SNR of NLoS and LoS.}
			\label{fig:sub1}
		\end{subfigure}
		\centering
		\begin{subfigure}{0.475\linewidth}
			\centering
			\includegraphics[width=0.97\linewidth]{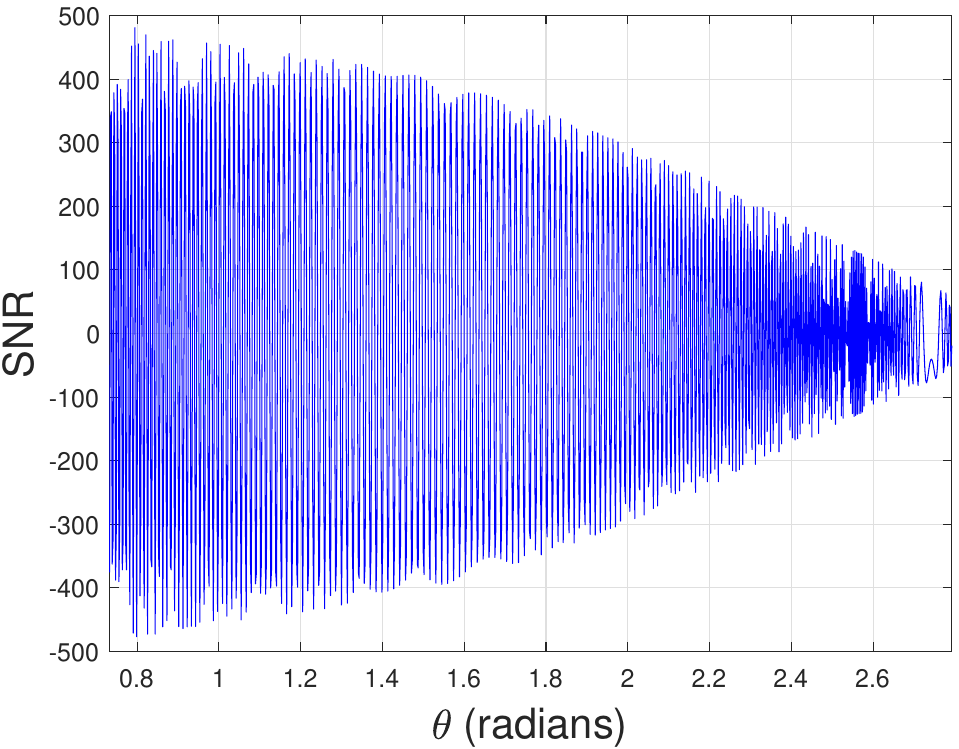}
			\caption{SNR of NLoS-LoS.}
			\label{fig:sub2}
		\end{subfigure}
		\\
		\centering
		\begin{subfigure}{0.475\linewidth}
			\centering
			\includegraphics[width=0.97\linewidth]{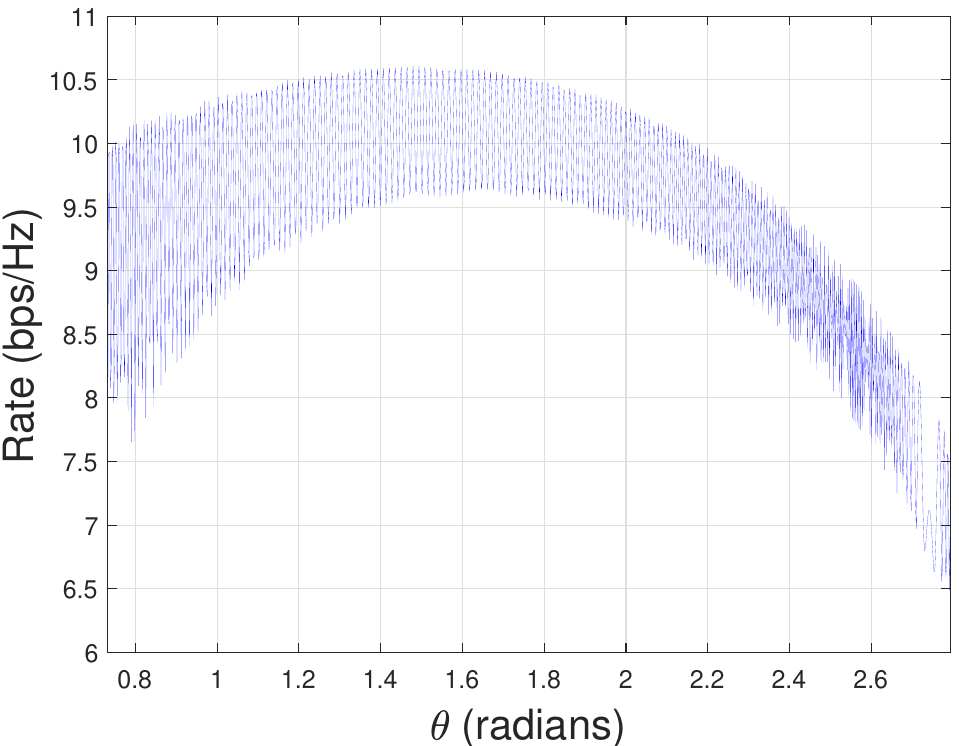}
			\caption{Rate v.s. $\theta$}
			\label{fig:sub3}
		\end{subfigure}
		\centering
		\begin{subfigure}{0.475\linewidth}
			\centering
			\includegraphics[width=0.97\linewidth]{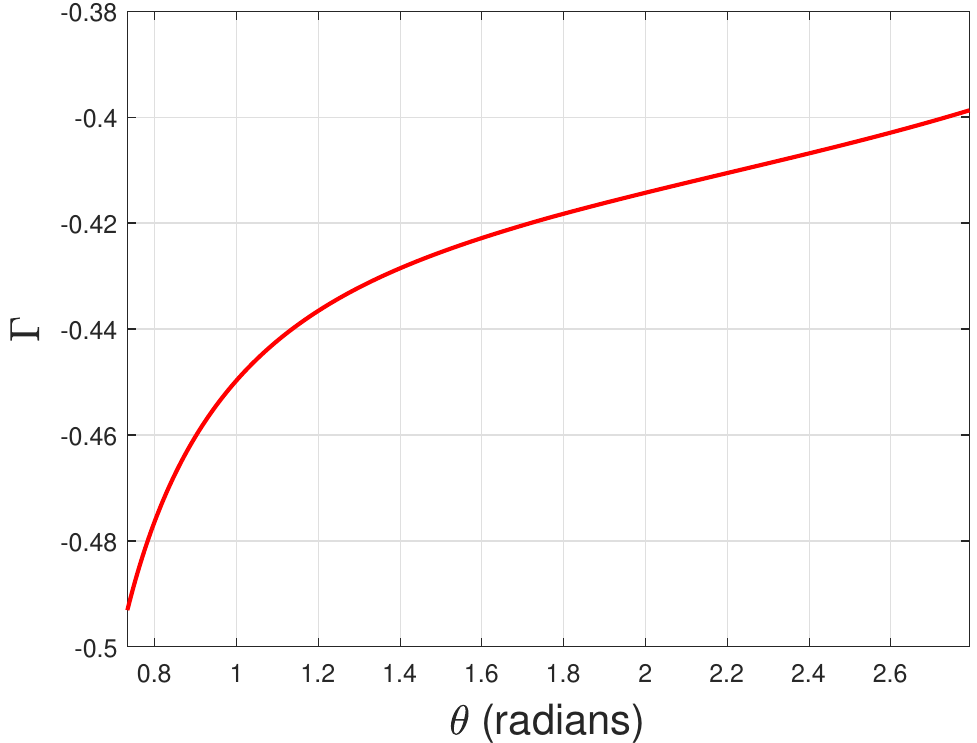}
			\caption{$\Gamma$ v.s. $\theta$}
			\label{fig:sub4}
		\end{subfigure}
		\caption{TE mode, $G_t=G_r=1, \sigma^2 = -90\,\text{dBm}, x_1=y_1=5\,\text{m}, x_0=0.5\,\text{m}, f = 5\,\text{GHz},\epsilon =  5.24$.}
		\label{fig:test}
	\end{figure}

	Despite depending on a single variable, $\text{SNR}_{2}(\theta)$ defies a closed-form solution due to its complex structure. The parameter $\theta$ appears within trigonometric functions in both numerator and denominator terms, making analytical simplification extraordinarily difficult. When applying numerical optimization methods like gradient descent, initial point selection becomes critical. As illustrated in Fig.~\ref{fig:test}, the highly oscillatory and irregular landscape of $\text{SNR}_{2}(\theta)$ hinders the identification of initial points that consistently converge to satisfactory local optima. Nevertheless, through rigorous examination of $\text{SNR}_{2}(\theta)$'s properties, we derive key insights enabling the development of a low-complexity closed-form solution.

	\begin{figure*}[t]
	\centering
	\includegraphics[width=4.75in]{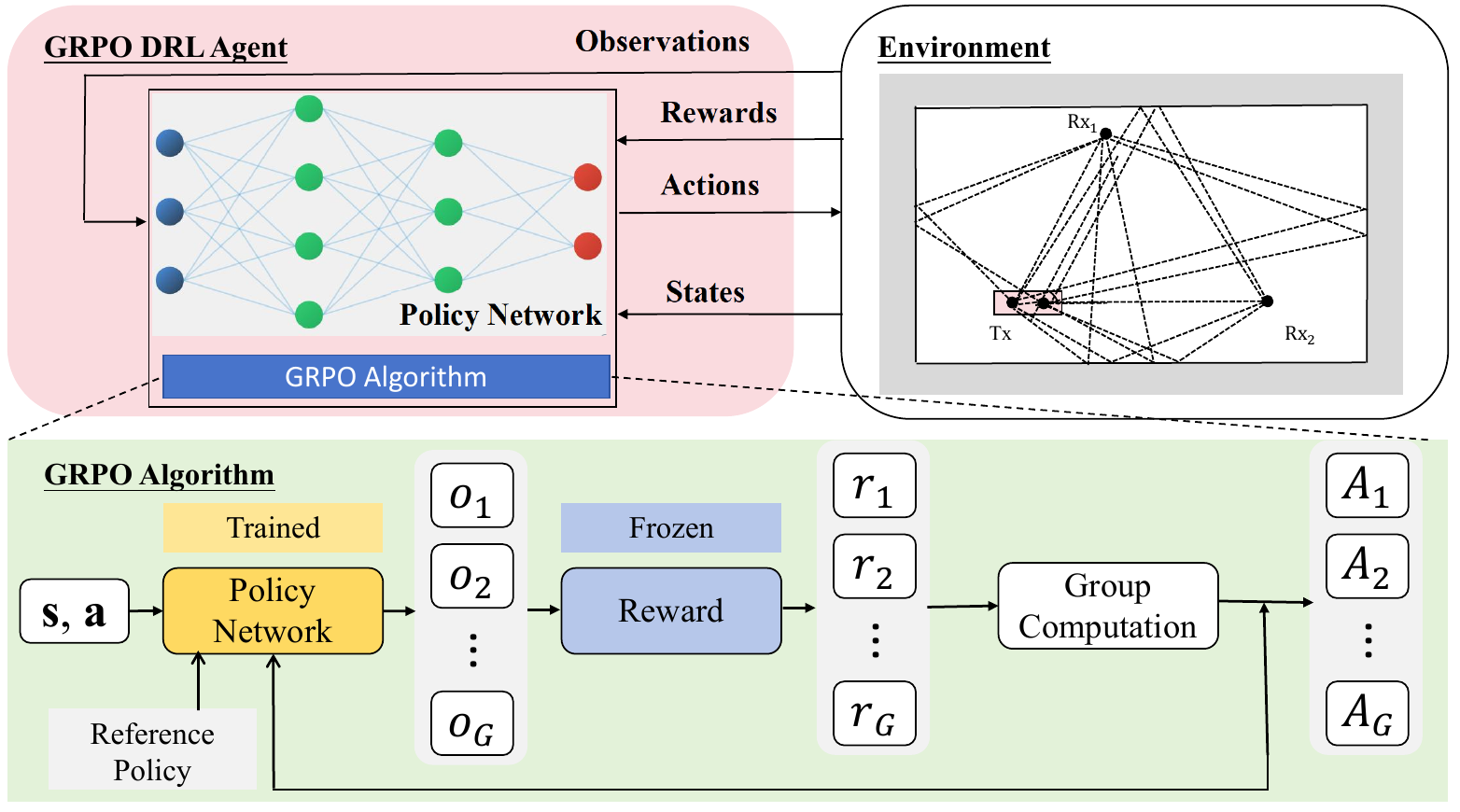}
	\caption{Illustration of the algorithm framework of GRPO training.}	\label{FGRPO}  
\end{figure*}

	To address Problem (P1) and obtain a closed-form solution, the properties of the function $\textsc{SNR}_{2}(\theta)$ play a critical role. To elucidate these properties by examples, we analyze a specialized case depicted in Fig. \ref{fig:test}. Two significant observations arise from Fig. \ref{fig:test}: \textit{1) LoS term within $\textsc{SNR}_{2}(\theta)$ predominantly governs the function, with NLoS term contributing only marginally; 2) the superposition of the LoS and NLoS terms yields a unimodal function, attaining its maximum at $\theta = \pi/2$ when this value is feasible. Moreover, the rate function exhibits oscillatory behavior synchronized with the frequency of the cosine term in the NLoS-LoS interaction component.} These findings provide the insights for devising a closed-form solution. 
	
	The proposed closed-form solution to Problem (P2), denoted as $\theta^*$, is systematically categorized into three cases, each expounded below on a case-by-case basis.
	
	\textbf{Case 1} (Left Region): When $\theta_l, \theta_r \in (\arctan \frac{y_1 - y_0}{x_1}, \frac{\pi}{2}]$ and $\theta_l \le \theta_r$, the solution lies the leftmost local maximizer to $\theta_r$ or $\theta_r$ itself. It can be seen that the oscillation comes from the $\cos$ function in the NLoS-LoS interaction term; and $(\sqrt{(y_1 - y_0)^2 + (2x_1 - \frac{y_1 - y_0}{\tan \theta})^2} - \frac{y_1 - y_0}{ \sin \theta})/\lambda$ is monotonically increasing in $\theta$ if $\theta \in [\arctan \frac{y_1 - y_0}{2x_1}, \pi)$. Through trigonometric identities and algebra manipulation, the
	closed-from solution $
	\theta^*$ is given in the below proposition.
	
	\begin{proposition}
		When $\theta_l, \theta_r \in (\arctan \frac{y_1 - y_0}{x_1}, \frac{\pi}{2})$ and $\theta_l \le \theta_r$, the closed-form solution $
		\theta^*$ to Problem (P2)  is given by
		\begin{equation}
			\theta^* = \arg\max \{\textsc{SNR}_{2}(f(n)),\textsc{SNR}_{2}(\theta_r)\},
		\end{equation}
		where if $\Gamma \ne 0$, then 
		\begin{eqnarray}
			f(n) = \arcsin \left( \frac{n\lambda(y_1 -y_0)}{\sqrt{(4x_1^2 - \frac{n^2\lambda^2}{4})^2 + (4x_1(y_1 -y_0))^2}} \right)  \nonumber \\
			- \arctan \left(\frac{-4x_1(y_1 -y_0)}{4x_1^2 - \frac{n^2\lambda^2}{4}}\right), \label{Proposition1} 
		\end{eqnarray} and
		\begin{equation}
		\!\!\!\!	n = \begin{cases}
				2 \lfloor\frac{\sqrt{(y_1 - y_0)^2 + (2x_1 - \frac{y_1 - y_0}{\tan \theta_r})^2} - \frac{y_1 - y_0}{ \sin \theta_r}}{\lambda}  \rfloor, & 0 < \Gamma, \label{Proposition11} \\ 
				2 \lfloor\frac{\sqrt{(y_1 - y_0)^2 + (2x_1 - \frac{y_1 - y_0}{\tan \theta_r})^2} - \frac{y_1 - y_0}{ \sin \theta_r}}{\lambda}  \rceil - 1, &  \Gamma < 0. \end{cases}
		\end{equation} 
	\end{proposition}
	
	\begin{IEEEproof}
		Please refer to Appendix C.
	\end{IEEEproof}
	
	\textbf{Case 2}  (Middle Region):  When $\theta_l \in (\arctan \frac{y_1 - y_0}{x_1}, \frac{\pi}{2})$ and $\theta_r \in [\frac{\pi}{2}, \pi - \arctan \frac{y_1 - y_0}{x_1})$, the solution lies the closest local maximizer to $\frac{\pi}{2}$ or $\frac{\pi}{2}$ itself.   $\theta^*$ is given by
	\begin{eqnarray}
	&& \!\!\!\!\!\!\!\!\!\!\!\!\!\!\!\!\!\!\!\!\!	\theta^* =  \nonumber \\ && \!\!\!\!\!\!\!\!\!\!\!\!\!\!\!\!\!\!\!\!\! \arg\max \{\textsc{SNR}_{2}(f(n_1)),  \textsc{SNR}_{2}(f(n_2)),  
		\textsc{SNR}_{2}(\pi/2)\}, 
	\end{eqnarray}
	where if $\Gamma \ne 0$, then $f(n_1)$ and $f(n_2)$ exist with 
	\begin{equation} 
		n_1 = \begin{cases}
			2\lfloor \frac{\sqrt{(y_1 - y_0)^2 +4x_1^2} - (y_1 -y_0)}{\lambda} \rfloor, & 0 < \Gamma, \\
			2\lfloor \frac{\sqrt{(y_1 - y_0)^2 +4x_1^2} - (y_1 -y_0)}{\lambda} \rceil - 1, &   \Gamma < 0,
		\end{cases}		
	\end{equation}
	and 
	\begin{equation}
		n_2 = \begin{cases}
			2\lceil \frac{\sqrt{(y_1 - y_0)^2 +4x_1^2} - (y_1 -y_0)}{\lambda} \rceil, & 0 < \Gamma, \\
			2\lfloor \frac{\sqrt{(y_1 - y_0)^2 +4x_1^2} - (y_1 -y_0)}{\lambda} \rceil +1, & \Gamma < 0.
		\end{cases}
	\end{equation}

	\textbf{Case 3} (Right Region): When $\theta_l, \theta_{r} \in [\frac{\pi}{2}, \pi - \arctan \frac{y_1 - y_0}{x_1})$ and $\theta_l \le \theta_r$, the solution lies the rightmost local maximizer to $\theta_l$ or $\theta_l$ itself. $\theta^*$ is given by
	\begin{equation}
		\theta^* = \arg\max \{\textsc{SNR}_{2}(f(n)),\textsc{SNR}_{2}(\theta_l)\},
	\end{equation}
	where  if $\Gamma \ne 0$, then $f(n)$ exists with 
	\begin{equation} 	
	\!\!\!\!	n = \begin{cases}
			2\lceil \frac{\sqrt{(y_1 - y_0)^2 + (2x_1 - \frac{y_1 - y_0}{\tan \theta_l})^2} - \frac{y_1 - y_0}{ \sin \theta_l}}{\lambda} \rceil, & 0 < \Gamma, \\
			2\lfloor \frac{\sqrt{(y_1 - y_0)^2 + (2x_1 - \frac{y_1 - y_0}{\tan \theta_l})^2} - \frac{y_1 - y_0}{ \sin \theta_l}}{\lambda} \rceil + 1, & \Gamma < 0.
		\end{cases}
	\end{equation}	 
The accuracy of the proposed closed-form solution will be validated in simulations.

\section{Group Relative Policy Optimization}\label{sec:GRPO}

Due to the complicated structure of Problem (P1), we customize the use of an advanced DRL algorithm, GRPO \cite{GRPO}, recently proposed by \textsc{DeepSeek}, offering an efficient approach to tackling this complex problem.

\subsection{MDP Definition}

As a foundation of DRL, we define Markov Decision Process (MDP) \textit{state}, \textit{action}, and \textit{reward} as follows:
\begin{itemize}
	\item \textit{MDP state} $\textbf{s}$ is defined by $(\mathcal{L},\mathcal{X},\{\textbf{h}_k({\bm{\theta}})\})$. Note that the state will transition based on the action taken.
	\item \textit{MDP action} $\textbf{a}$ is defined by $({\bm{\theta}},\{\bm{\omega}_k\})$. The action is the optimization variable for Problem (P1).
	\item  \textit{MDP reward} $r$ is defined by  
	\begin{equation}
		r = \sum_{k=1}^K \log_2\left(1 + \frac{|\textbf{h}_k^H(\bm{\theta}) \bm{\omega}_k|^2}{\sum_{k' \ne k} |\textbf{h}_{k'}^H(\bm{\theta})\bm{\omega}_k|^2 + \sigma^2}\right),
	\end{equation}
	if constraints \eqref{24a}-\eqref{24c} are satisfied. 
	Otherwise, the \textit{MDP reward} $r$ will be a large negative number. The reward is the optimization objective for Problem (P1).
	\item \textit{MDP observation} $\textbf{o}$ corresponds to the successor state following an action.
\end{itemize}
Based on the above MDP definition, the optimal action made from states leads to an optimized solution to Problem (P1).

\subsection{GRPO Solution}
DRL centers on the MDP framework, where an agent learns an optimal policy by maximizing cumulative reward, guided by state-action value functions approximated with neural networks.  PPO stands as a prominent policy gradient algorithm in DRL that addresses the instability of traditional policy optimization methods through a constrained update mechanism. Introduced by OpenAI in 2017 \cite{PPO}, PPO's core innovation lies in its clipped surrogate objective function, which restricts policy changes by limiting the ratio between new and old policy probabilities to a predefined interval, effectively creating a trust region that prevents destructively large updates. 

GRPO is a nontrivial variant of PPO \cite{PPO}. Unlike PPO,  GRPO offers the distinct benefit of reduced resource demands during training, as noted in \cite{GRPO}. Specifically, GRPO replaces the resource-intensive critic network of PPO with a set of trajectories, collectively termed a ``group''. The advantage function for a policy is therefore calculated using relative advantage functions within the group. This advantage is then used to update the actor network through stochastic gradient descent. This framework is shown in Fig. \ref{FGRPO}.

According to \cite{GRPO}, GRPO aims to optimize the parameters $w$ by maximizing the objective $\mathbb{E}\{J_\text{GRPO}(w)\}$, where $J_\text{GRPO}(w)$ is defined in \eqref{GRPO}.
\begin{figure*}
	\begin{equation}
		J_\text{GRPO}(w) =  \frac{1}{G} \sum_{g=1}^G \frac{1}{O_g} \sum_{t=1}^{O_g} \left\{ 
		\min \left(
		\frac{\pi_w(\textbf{o}_{g,t}|\textbf{a},\textbf{o}_{g,<t})}{\pi_{w_\text{old}}(\textbf{o}_{g,t}|\textbf{a},\textbf{o}_{g,<t})}
		, \text{clip}\left(\frac{\pi_w(\textbf{o}_{g,t}|\textbf{a},\textbf{o}_{g,<t})}{\pi_{w_\text{old}}(\textbf{o}_{g,t}|\textbf{a},\textbf{o}_{g,<t})},1-c,1+c\right)\right)\widehat{A}_{g} - \eta \mathbb{D}_\text{KL}[\pi_{w}\|\pi_\text{ref}]
		\right\}  \label{GRPO}
	\end{equation}
	\hrule
\end{figure*}
As GRPO's main innovation, the advantage function in \eqref{GRPO} is computed through group relative advantage estimation, given by
\begin{equation}
	\widehat{A}_{g} = \frac{r_g - \text{mean}(\{r_1,\ldots,r_G\})}{\text{std}(\{r_1,\ldots,r_G\})}, \label{27} 
\end{equation}
where operations $\text{mean}\{\cdot\}$ and $\text{std}\{\cdot\}$ denote the mean and standard deviation of set $\{\cdot\}$, respectively.

Specifically, we only have a terminal reward, where a non-zero reward is only provided at the final step of an episode. This is particularly fitting for our problem, since the solution of Problem (P2) is given at the last step.
According to \cite{GRPO}, the clip function in \eqref{GRPO} is used for limiting drastic policy updates, given by
\begin{eqnarray}
	&& \!\!\!\!\!\!\!\!\!\!\!\! \text{clip}\left(\frac{\pi_w(\textbf{o}_{g,t}|\textbf{a},\textbf{o}_{g,<t})}{\pi_{w_\text{old}}(\textbf{o}_{g,t}|\textbf{a},\textbf{o}_{g,<t})},1-c,1+c\right) \nonumber \\
	&&\!\!\!\!\!\!\!\!\!\!\!\!  = \max\left(\min\left(\frac{\pi_w(\textbf{o}_{g,t}|\textbf{a},\textbf{o}_{g,<t})}{\pi_{w_\text{old}}(\textbf{o}_{g,t}|\textbf{a},\textbf{o}_{g,<t})},1+c\right),1-c\right),
\end{eqnarray}
where $c$ denotes the clipping threshold, and $\pi_{w_\text{old}} = \pi_\text{ref}$ for the very start of GRPO training. 

Kullback-Leibler (KL) divergence in \eqref{GRPO} is utilized to regulate the difference between the updated policy and the reference policy $\pi_\text{ref}$, with its importance controlled by KL penalty $\eta \geq 0$. According to \cite{GRPO}, KL divergence can be approximately calculated by 
\begin{eqnarray}
	    \mathbb{D}_\text{KL}[\pi_{w}\|\pi_\text{ref}] \approx \frac{\pi_\text{ref}(\textbf{o}_{g,t}|\textbf{a},\textbf{o}_{g,<t})}{\pi_w(\textbf{o}_{g,t}|\textbf{a},\textbf{o}_{g,<t})}  -  \nonumber \\
     \log \frac{\pi_\text{ref}(\textbf{o}_{g,t}|\textbf{a},\textbf{o}_{g,<t})}{\pi_w(\textbf{o}_{g,t}|\textbf{a},\textbf{o}_{g,<t})} -1.
\end{eqnarray}

Finally, the algorithm for GRPO training is given in Algorithm 2, where we invoke the PPO algorithm for $T$ steps to obtain a reference policy. Note that the steps of $T$ are small, and the PPO algorithm does not need to converge.

\begin{algorithm}[!t]
	\caption{Algorithm for GRPO Training}
	\label{algorithm:GRPO}
	\KwIn{$\mathcal{L},\,\mathcal{X}, T$ and other hyperparameters}
	Generate the reference policy $\pi_\text{ref}$ by training PPO for $T$ steps \\ 
	\For{Iteration =$1,\ldots,I$}{
		Update the old policy $\pi_{w_\text{old}} \leftarrow \pi_w$ \\
		Sample $G$ outputs $\{\textbf{o}_g\}_{g=1}^G \sim \pi_{w_\text{old}}(\cdot|\textbf{a})$ for a randomly selected action $\textbf{a}$  \\
		Compute rewards $\{r_g\}_{g=1}^G$ for each $\textbf{o}_g,\,g \in [G]$ \\
		Compute advantage function $\widehat{A}_{g}$ using \eqref{27} \\ 
		\For{GRPO iteration = $1,\ldots,\mu$}{
			Update the policy $\pi_w$ through maximizing \eqref{GRPO}
		}		 
	}
	\KwOut{Optimized policy $\pi_{w}$}
\end{algorithm}

\subsection{Computational Complexity}

To compare the complexity of GRPO and PPO, we analyze the computational complexity of the only actor network (GRPO) and the actor and critic networks (PPO). Suppose that both actor and critic networks have $L_\text{hidden}$ layers and each layer has $L$ neurons. According to \cite{judd1990neural}, the computational complexity for the actor network is on the order of $\mathcal{O}(d_s L + \underbrace{L^2 + \cdots + L^2}_{L_\text{hidden}} + L d_a) = \mathcal{O}(L(d_s + d_a + L L_\text{hidden})) $, and critic network is on the order of $\mathcal{O}(d_s   L + \underbrace{L^2 + \cdots + L^2}_{L_\text{hidden}} + L) = \mathcal{O}(L(d_s+1+L L_\text{hidden}))$, where $d_s$ denotes the state dimension and is equal to $2K + 2 + 2M$ and $d_a$ denotes the action dimension and is equal to $N + 2KN$. Therefore, PPO and GRPO have $\mathcal{O}(L(2d_s+d_a+1+2L L_\text{hidden}))$ and $\mathcal{O}(L(d_s + d_a + L L_\text{hidden}))$ computational complexity for neural networks, respectively. It can be seen that GRPO reduces the computational complexity of neural networks by approximately half compared to PPO. 
 
\section{Simulations}\label{sec:result}
\subsection{Settings and Baselines}
Simulation settings are assigned as follows: AWGN variance is $\sigma^2 = -90$ dBm, and the maximal transmit power is $P_{\max}=1$ W.  The relative permittivity of walls is $\epsilon = 5.24$, which is that for concrete \cite{hoydis2022sionna}.   The antenna gains are set as $G_t = G_r = 1$. The neural network configurations in PPO and GRPO are listed in Table III, where we adopt mmoderate-scale actor and critic networks for our problem. Baseline algorithms for comparison with GRPO are below. 
 
 \textbf{PPO Init.} (\textit{PPO \cite{PPO} training $T=25$ K steps as our GRPO reference policy}): The baseline consists of training PPO \cite{PPO} for 25 K steps, a process that takes roughly $3$ minutes on our computer. Our GRPO method uses this trained policy as its reference policy for initialization.
 
  \begin{table}[t]
	\centering
	\caption{\textsc{Configuration { \& Learning Parameters}}}
	\label{tab:optimized}
	\begin{tabularx}{0.95\columnwidth}{@{} X X X X @{}}
		\toprule
		\multicolumn{4}{@{}c}{\textbf{Actor Network (GRPO, PPO, PPO Init.)}} \\
		\textbf{Component} & \textbf{Type} & \textbf{Input/Output} & \textbf{Activation}\\
		Input Layer       & Linear      & Input Size$\times$256    & ReLU\\ 
		Hidden Layer      & Linear      & 256$\times$256           & ReLU \\     
		Output Layer      & Linear      & 256$\times$Output Size   & --     \\
		\multicolumn{4}{@{}c}{\textbf{Critic Network  (PPO, PPO Init.)}} \\
		\textbf{Component} & \textbf{Type} & \textbf{Input/Output} & \textbf{Activation}\\
		Input Layer      & Linear      & Input Size$\times$256          & ReLU      \\         
		Hidden Layer    & Linear      & 256$\times$256             & ReLU      \\       
		Output Layer    & Linear      & 256$\times$Output Size     & --           \\    
          \multicolumn{4}{@{}c}{{\textbf{Learning Parameters  (GRPO, PPO, PPO Init.)}}} \\
{\textbf{Learning}} & {\textbf{KL}} & {\textbf{Clipping}} & {\textbf{Batch}} \\
{\textbf{Rate}} & {\textbf{Penalty}} & {\textbf{Threshold}} & {\textbf{Size}} \\
{$9.46\times 10^{-4}$} & {$0.0001$} & {$0.1$} & {$64$}\\    
\hline
	\end{tabularx} 
\end{table}

\begin{figure*}[t]
	\centering
	\begin{subfigure}{0.24\linewidth}
		\centering
		\includegraphics[width=0.98\linewidth]{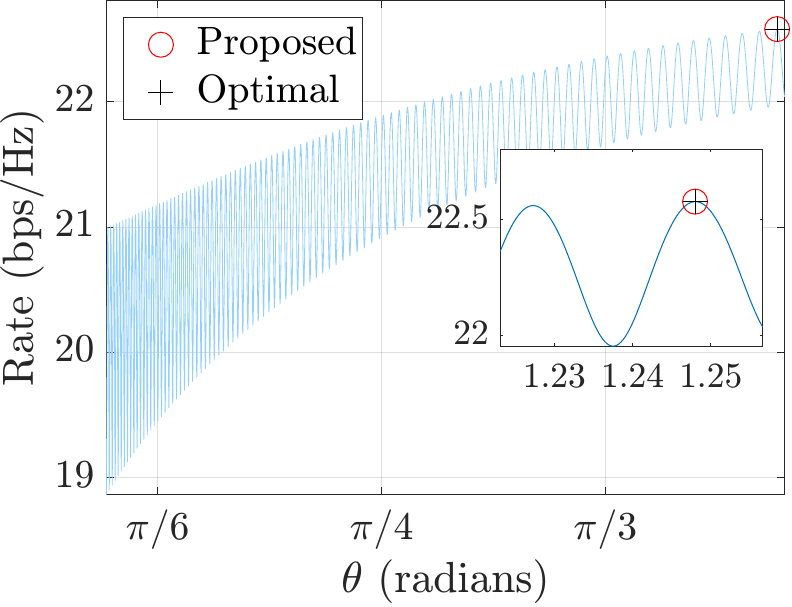}
		\caption{$0.23 \pi \le \theta \le 0.4\pi$, TM}
		\label{7a}
	\end{subfigure}
	\hfill
	\begin{subfigure}{0.24\linewidth}
		\centering
		\includegraphics[width=0.98\linewidth]{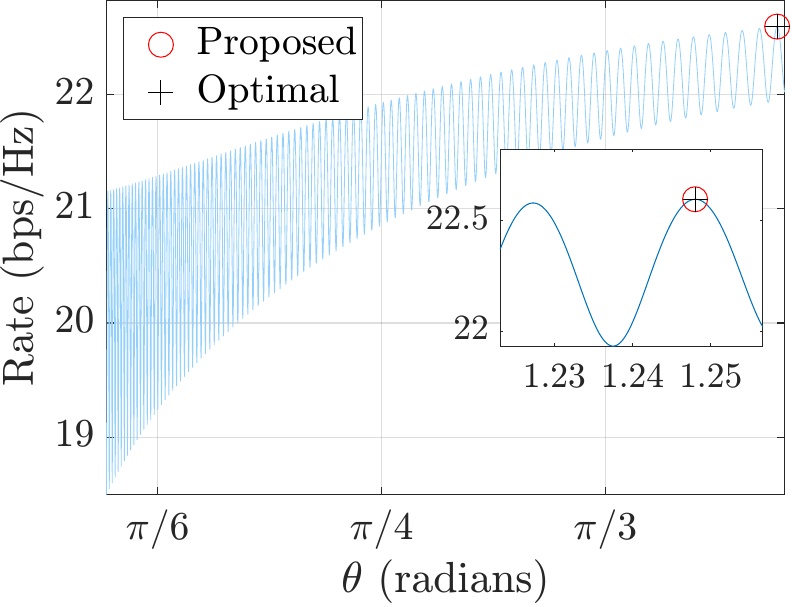}
		\caption{$0.23 \pi \le \theta \le 0.4\pi$, TE}
		\label{7b}
	\end{subfigure}
	\hfill
	\begin{subfigure}{0.24\linewidth}
		\centering
		\includegraphics[width=0.98\linewidth]{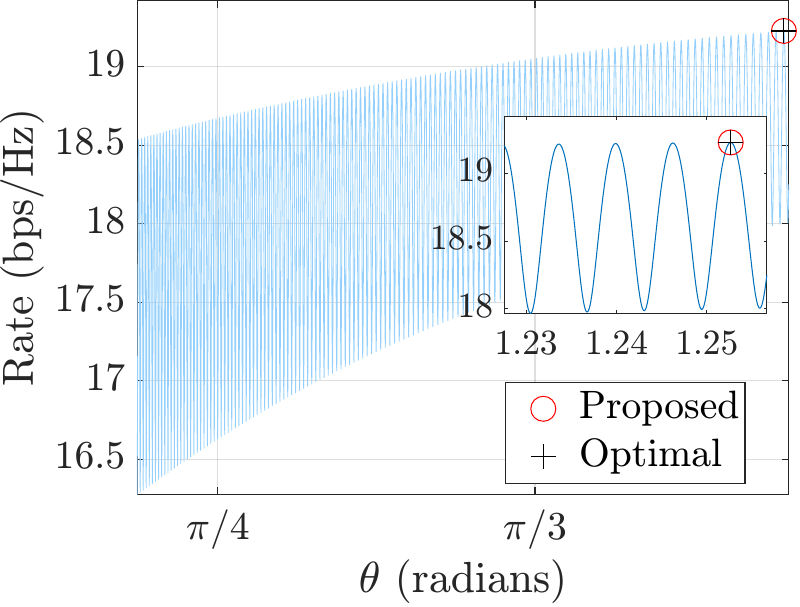}
		\caption{$0.23 \pi \le \theta \le 0.4\pi$, TM}
		\label{fig:sub11}
	\end{subfigure}
	\hfill
	\begin{subfigure}{0.24\linewidth}
		\centering
		\includegraphics[width=0.98\linewidth]{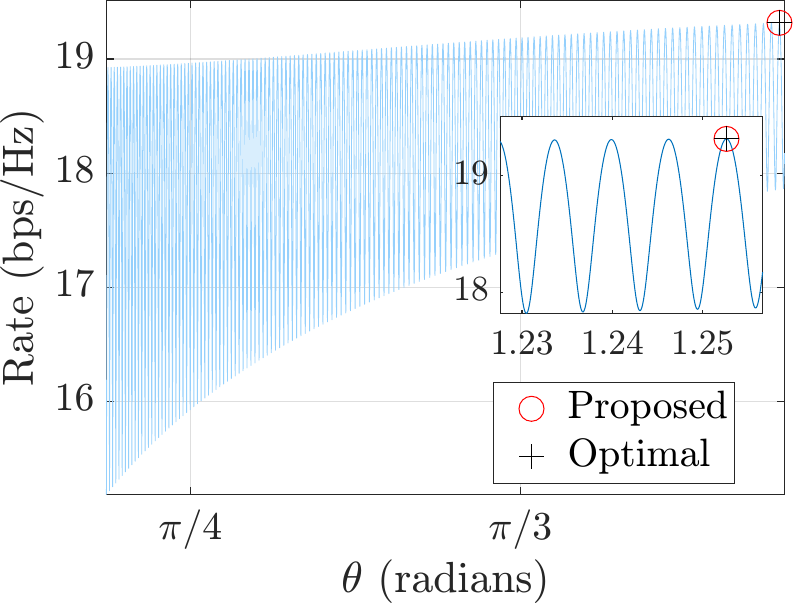}
		\caption{$0.23 \pi \le \theta \le 0.4\pi$, TE}
		\label{fig:sub44}
	\end{subfigure}
	
	\begin{subfigure}{0.24\linewidth}
		\centering
		\includegraphics[width=0.98\linewidth]{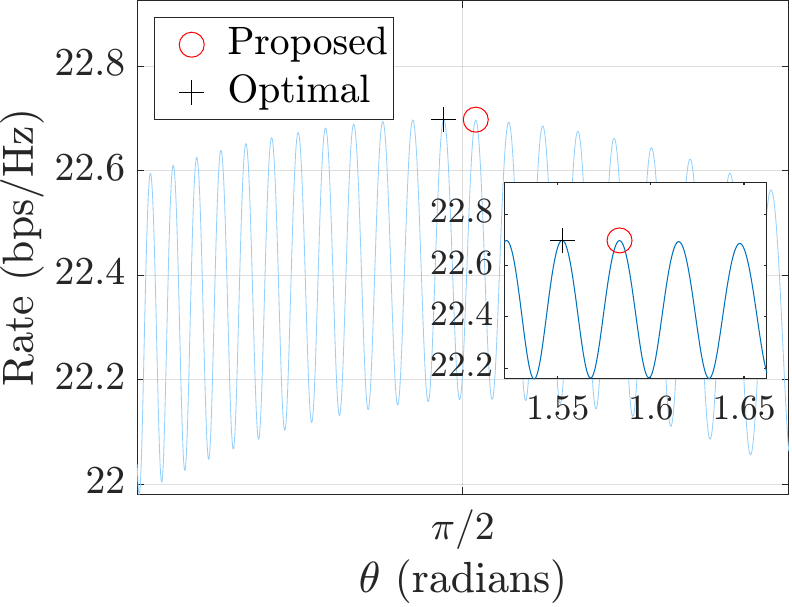}
		\caption{$0.4\pi \le \theta \le 0.6\pi$, TM}
		\label{7c}
	\end{subfigure}
	\hfill
	\begin{subfigure}{0.24\linewidth}
		\centering
		\includegraphics[width=0.98\linewidth]{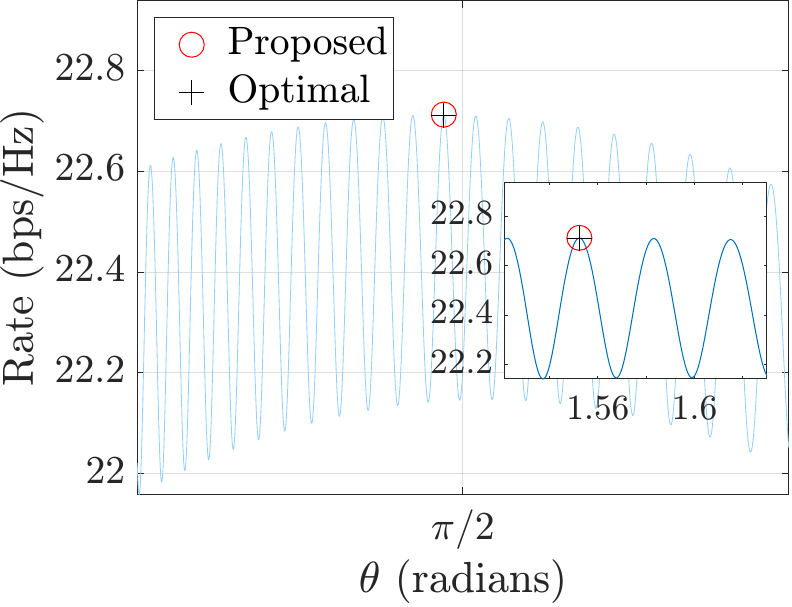}
		\caption{$0.4\pi \le \theta \le 0.6\pi$, TE}
		\label{7d}
	\end{subfigure}
	\hfill
	\begin{subfigure}{0.24\linewidth}
		\centering
		\includegraphics[width=0.98\linewidth]{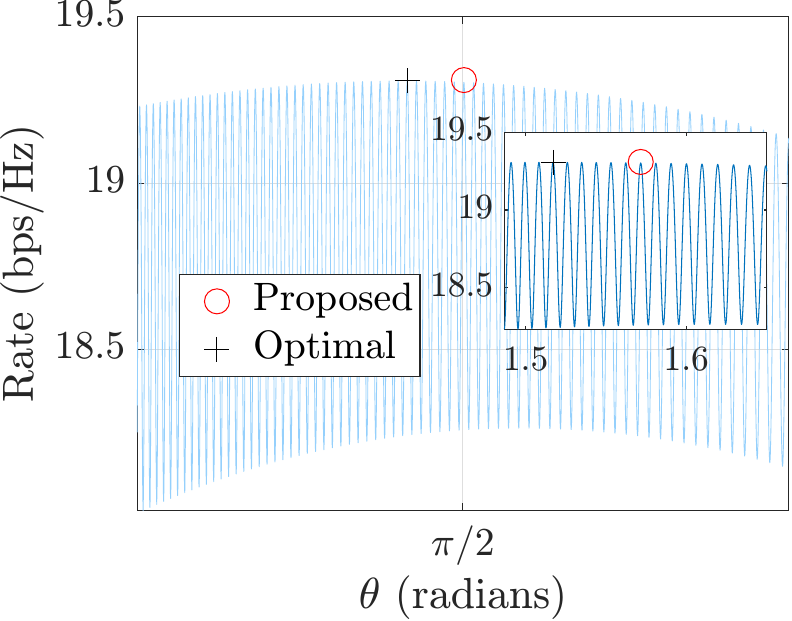}
		\caption{$0.4\pi \le \theta \le 0.6\pi$, TM}
		\label{fig:sub22}
	\end{subfigure}
	\hfill
	\begin{subfigure}{0.24\linewidth}
		\centering
		\includegraphics[width=0.98\linewidth]{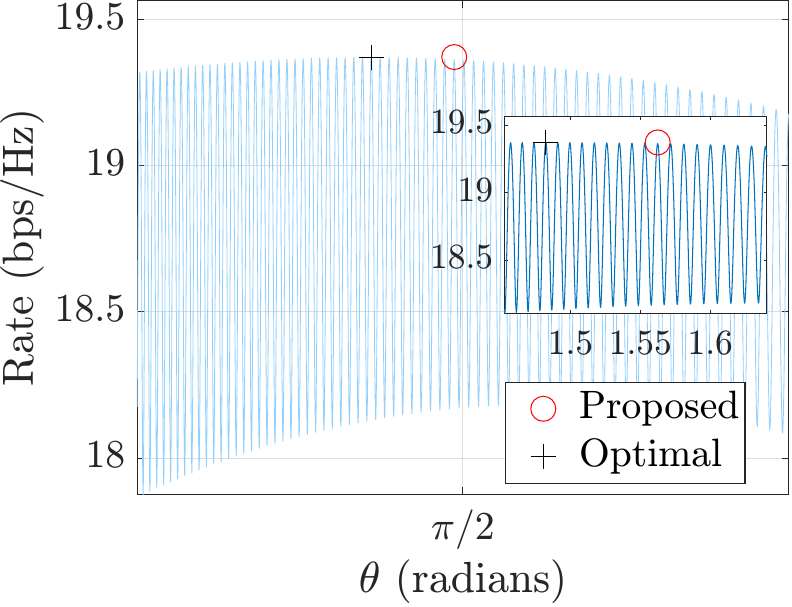}
		\caption{$0.4\pi \le \theta \le 0.6\pi$, TE}
		\label{fig:sub55}
	\end{subfigure}
	
	\begin{subfigure}{0.24\linewidth}
		\centering
		\includegraphics[width=0.98\linewidth]{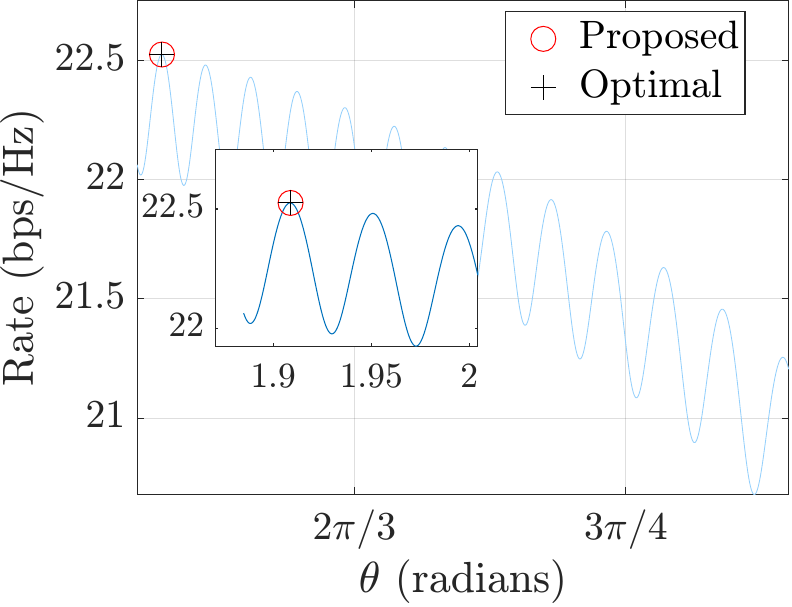}
		\caption{$0.6\pi \le \theta \le 0.82\pi$, TM}
		\label{7e}
	\end{subfigure}
	\hfill
	\begin{subfigure}{0.24\linewidth}
		\centering
		\includegraphics[width=0.98\linewidth]{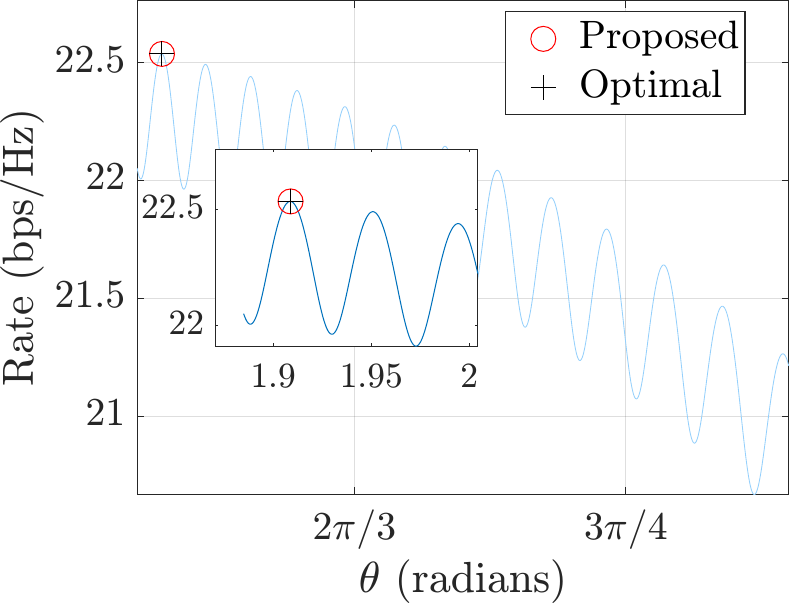}
		\caption{$0.6\pi \le \theta \le 0.82\pi$, TE}
		\label{7f}   
	\end{subfigure}
	\hfill
	\begin{subfigure}{0.24\linewidth}
		\centering
		\includegraphics[width=0.98\linewidth]{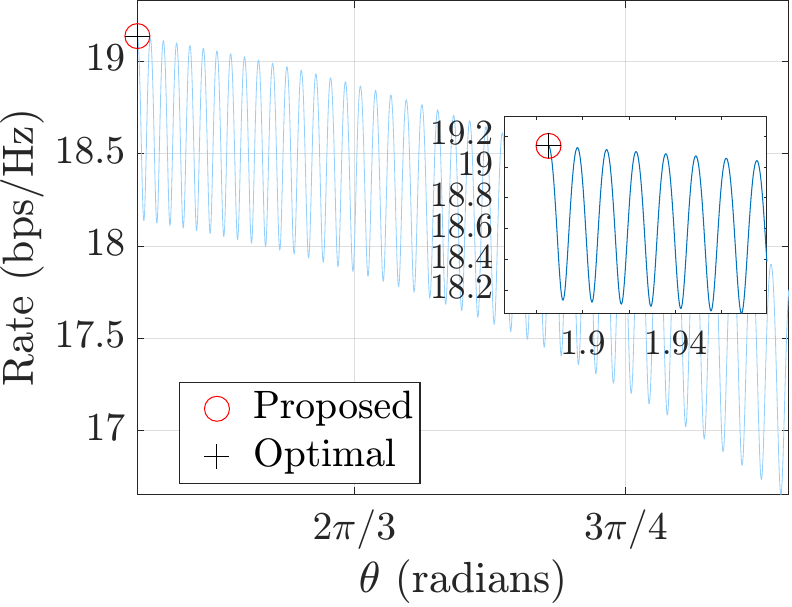}
		\caption{$0.6\pi \le \theta \le 0.82\pi$, TM}
		\label{fig:sub33}
	\end{subfigure}
	\hfill
	\begin{subfigure}{0.24\linewidth}
		\centering
		\includegraphics[width=0.98\linewidth]{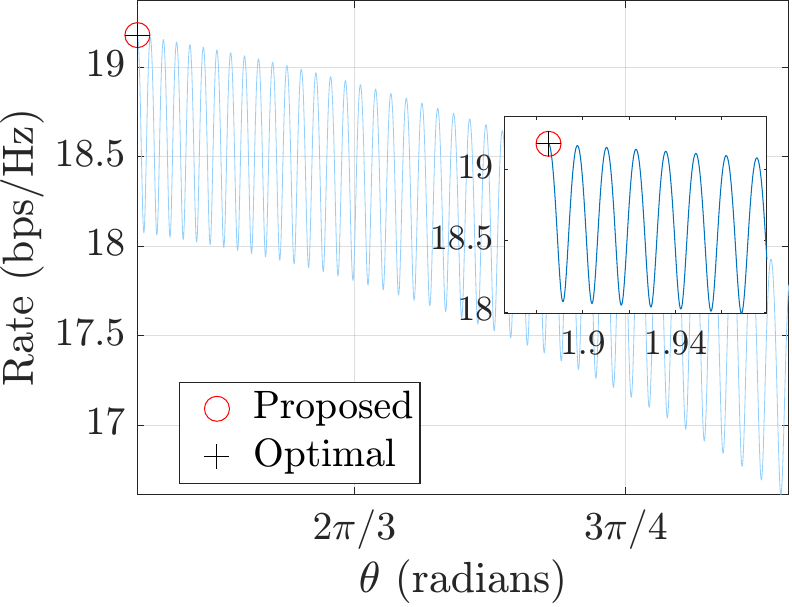}
		\caption{$0.6\pi \le \theta \le 0.82\pi$, TE}
		\label{fig:sub66}         
	\end{subfigure}	
	
	\caption{$G_t=G_r=1, \sigma^2 = -90\,\text{dBm},  f = 5\,\text{GHz}, \epsilon = 5.24$m, where $x_1=y_1=4\text{m}$,  $y_0=2\text{m}$ for \eqref{7a},\eqref{7b},\eqref{7c},\eqref{7d},\eqref{7e},\eqref{7f}, and  $x_1=y_1=4\text{m}, y_0=0.5\text{m}$ for \eqref{fig:sub11},\eqref{fig:sub44},\eqref{fig:sub22},\eqref{fig:sub55},\eqref{fig:sub33},\eqref{fig:sub66}.}
	\label{7}
\end{figure*}

 \textbf{PPO} (\textit{PPO in \cite{PPO}}): This baseline trains PPO \cite{PPO} until $3$ million steps. It is known that PPO has a superior performance. This baseline is to test how much the performance gain can be obtained from PPO in \cite{PPO}.

\textbf{A2C} (\textit{A2C applied in \cite{Chao}}): This baseline trains A2C until $3$ million steps. A2C algorithm combines policy learning with value estimation to reduce variance while maintaining direct policy optimization. Differently, A2C lacks PPO's trust region constraints, making it simpler but less stable during training. A2C shows its success in FAS-ISAC \cite{Chao}. 

 \textbf{WMMSE-FP} (\textit{Fixed fluid antenna position, WMMSE beamforming and power allocation \cite{WMMSE}}): This baseline fixes the fluid antenna position. Meanwhile, WMMSE \cite{WMMSE} is applied to optimize beamforming and transmit power. We randomly sample $200$ antenna positions for averaging its performance, since it takes a fixed antenna position and is susceptible to unsatisfactory positions.  

 {
 \textbf{WMMSE-GS} (\textit{Fluid antenna position with grid search (GS),  beamforming, and power allocation with  WMMSE \cite{WMMSE}}): Search fluid antenna positions via grid. The grid is given by ${0, \Delta, 2\Delta, \ldots, 25\Delta}$, where $\Delta = \lambda/2 = 0.03$m is the minimal  distance between two adjacent antennas. This baseline exhaustively evaluates all $\binom{26}{N}$ possible antenna positions to find the optimal combination. For each combination, beamforming and power allocation are optimized using WMMSE algorithm \cite{WMMSE}. This baseline aims to examine the GRPO advantage over the state-of-the-art numerical optimization method.
 }

\begin{table}[t]
	\renewcommand\arraystretch{1.3}
	\centering
	\captionof{table}{Performance comparison of PPO Init. and the closed-form in Section IV. The layout is rectangular with corner points $\{(0,0), (0,5), (5,5), (5,0)\}$ (Unit: m). Fluid antenna position is $y = 0.5$m, $0.6\pi \le \theta \le 0.82\pi$ (for Rx position 1), and $0.23\pi \le \theta \le 0.3\pi$ (for Rx position 2).}
	\label{tab:methods_scenarios}
	\begin{tabular}{l|c|c|c}
	\hline
	\multicolumn{1}{c|}{\textbf{Frequency}, \textbf{Rx Location}} & \textbf{Method} & \textbf{\begin{tabular}[c]{@{}c@{}}Rate\\ (bps/Hz)\end{tabular}} & {\textbf{Solution} $\theta^*$} \\ \hline
	\multirow{2}{*}{{\begin{tabular}[c]{@{}c@{}}5GHz, RX Position 1\\ $(1.5\text{m}, 1.5\text{m})$\end{tabular}}}  & {PPO Init.}  & 25.4154  & 0.6084$\pi$ \\ \cline{2-4}
	& {Closed-Form} & \textbf{25.4155}  & 0.6084$\pi$ \\ \hline
	\multirow{2}{*}{{\begin{tabular}[c]{@{}c@{}}60GHz, RX Position 1\\ $(1.5\text{m}, 1.5\text{m})$\end{tabular}}} & {PPO Init.}    & 18.8406  & 0.6010$\pi$ \\ \cline{2-4}
	 & {Closed-Form}    & \textbf{18.8406}  & 0.6010$\pi$ \\ \hline
	\multirow{2}{*}{{\begin{tabular}[c]{@{}c@{}}5GHz, RX Position 2\\ $(4.25\text{m}, 3\text{m})$\end{tabular}}} & {PPO Init.}    & 19.9056  & 0.2967$\pi$ \\ \cline{2-4}
	 & {Closed-Form}    & \textbf{20.9301}  & 0.2998$\pi$ \\ \hline
	\multirow{2}{*}{{\begin{tabular}[c]{@{}c@{}}60GHz, RX Position 2\\ $(4.25\text{m}, 3\text{m})$\end{tabular}}} & {PPO Init.}    & 14.4242  & 0.2967$\pi$ \\ \cline{2-4}
	 & {Closed-Form}    & \textbf{16.0837}  & 0.2998$\pi$ \\ \hline 
	\end{tabular}
\end{table}

\begin{figure*}[t]	 
	\centering
	\begin{subfigure}{0.49\linewidth}
		\centering
		\includegraphics[width=1\linewidth]{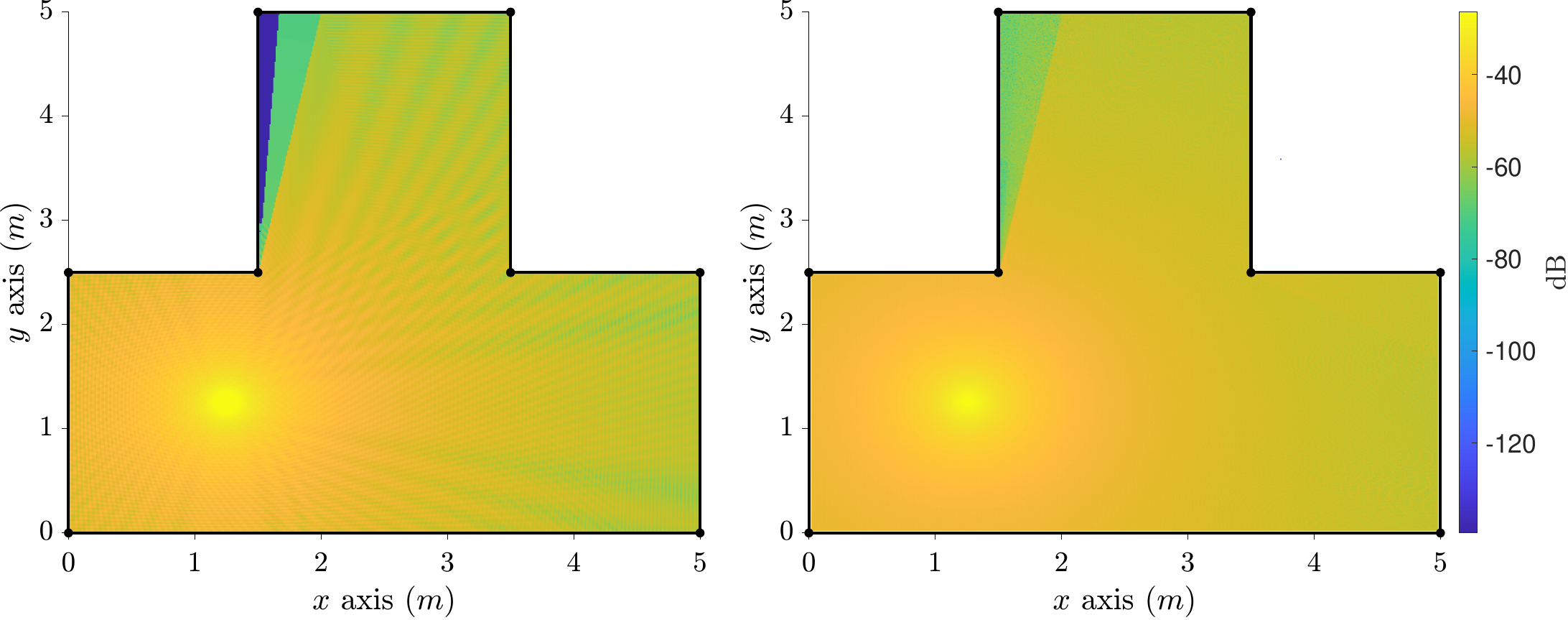}
		\caption{Layout 1, $f$ = 5 GHz}
		\label{fig:sub222}
	\end{subfigure}
	\centering
	\begin{subfigure}{0.49\linewidth}
		\centering
		\includegraphics[width=1\linewidth]{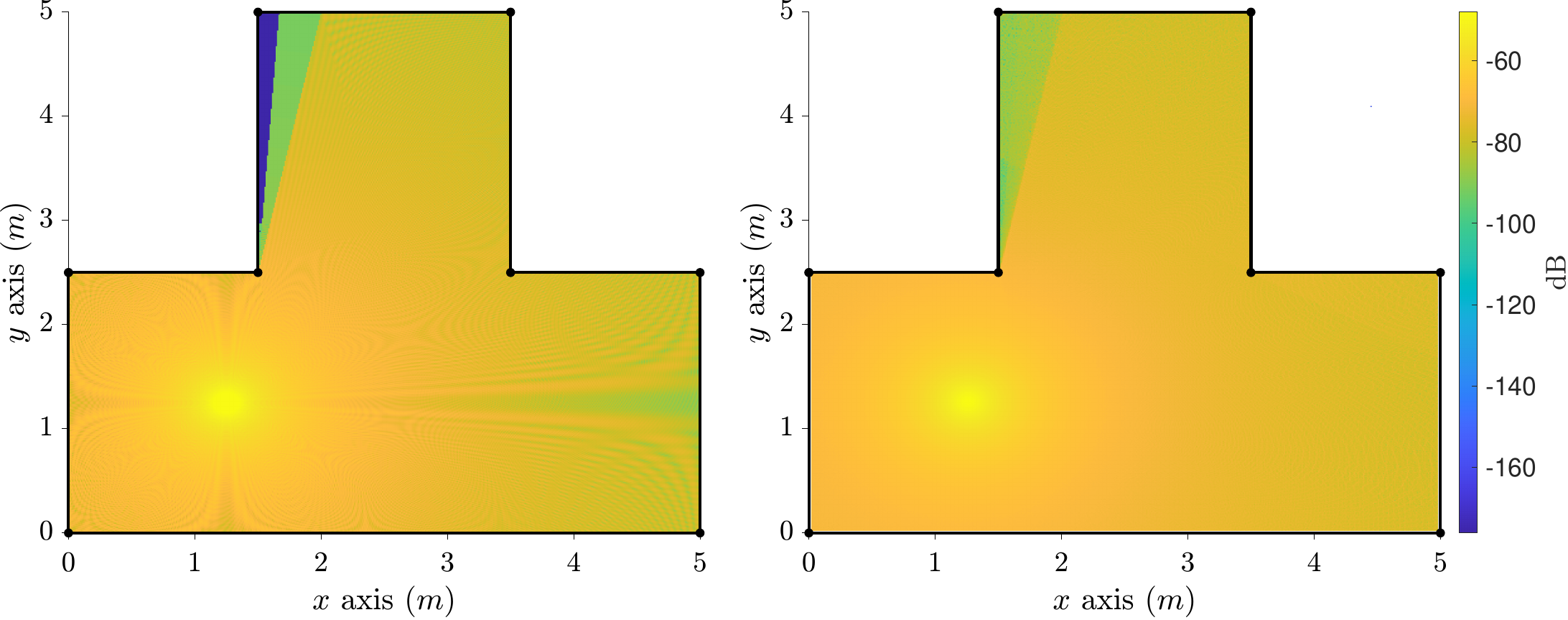}
		\caption{Layout 1, $f$ = 60 GHz}
		\label{fig:sub22222}
	\end{subfigure}
	\centering
	\begin{subfigure}{0.49\linewidth}
		\centering
		\includegraphics[width=1\linewidth]{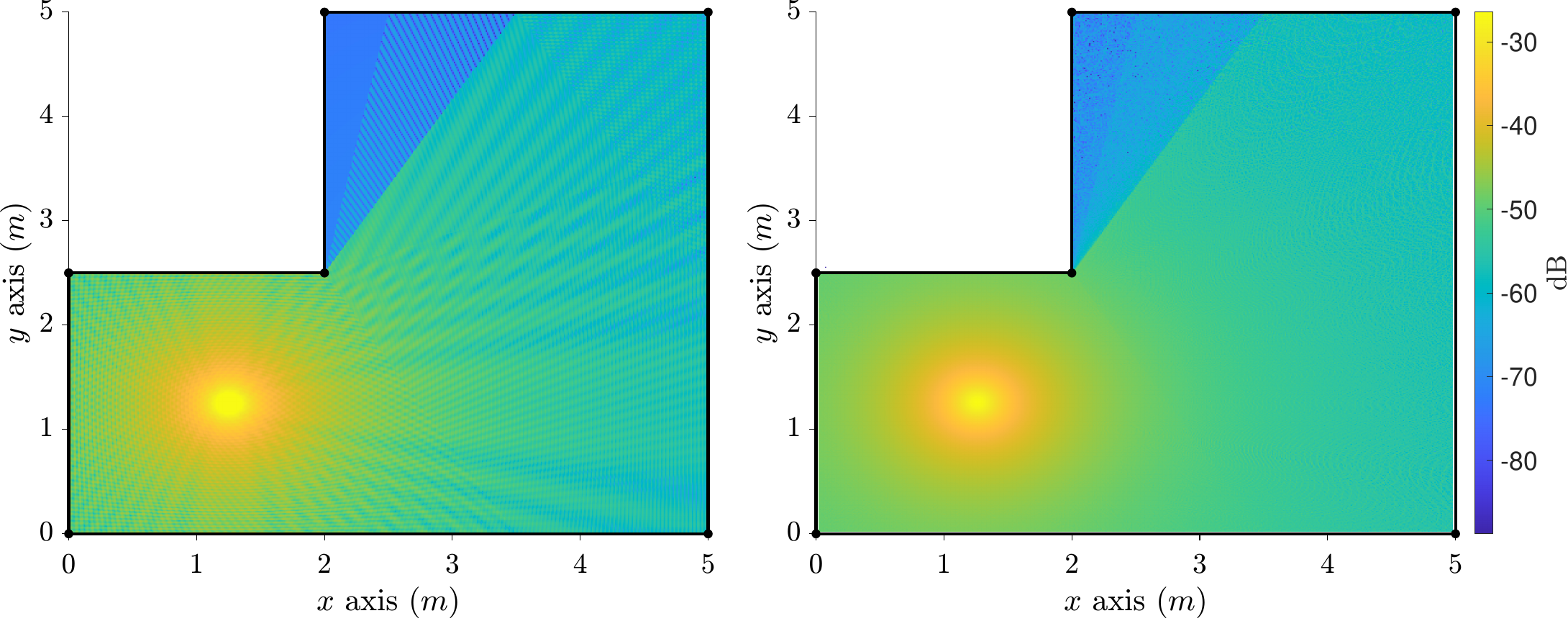}
		\caption{Layout 2, $f$ = 5 GHz}
		\label{fig:sub555}
	\end{subfigure} 
	\centering
	\begin{subfigure}{0.49\linewidth}
		\centering
		\includegraphics[width=1\linewidth]{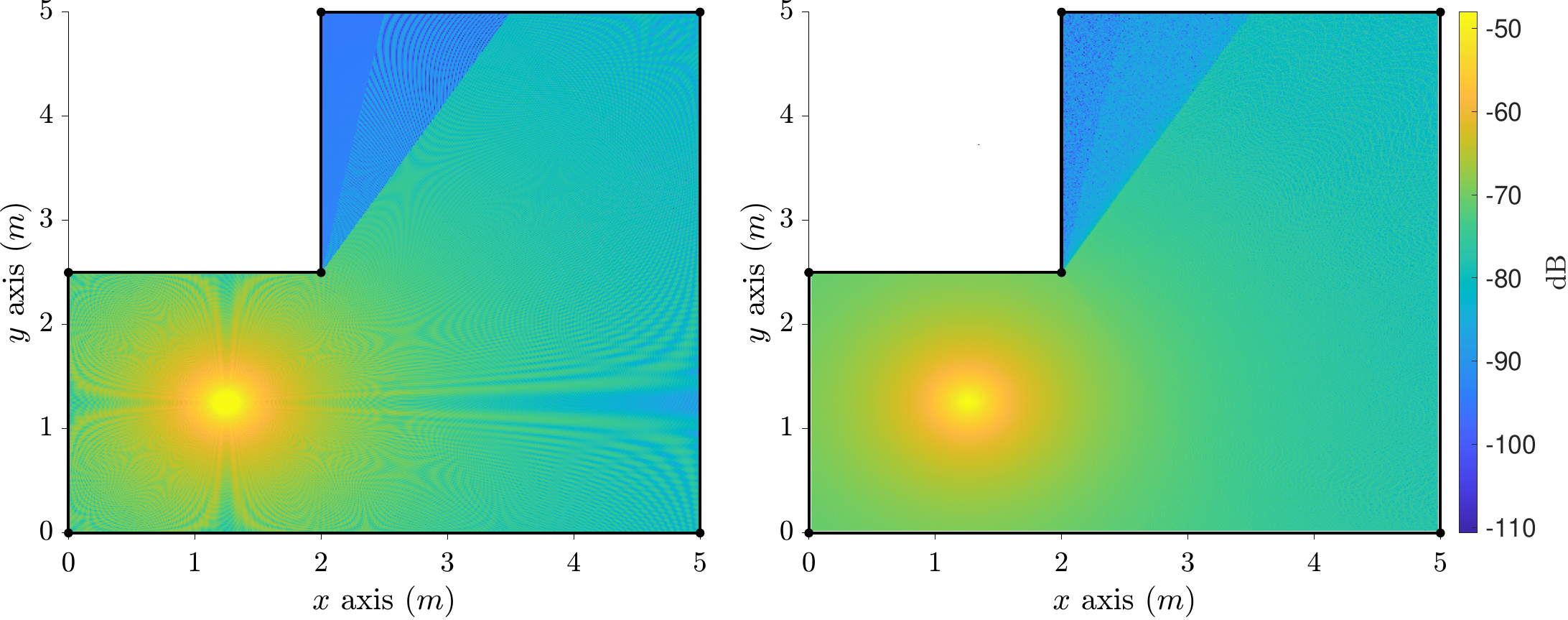}
		\caption{Layout 2, $f$ = 60 GHz}
		\label{fig:sub5555}
	\end{subfigure} 
	\begin{subfigure}{0.49\linewidth}
		\centering
		\includegraphics[width=1\linewidth]{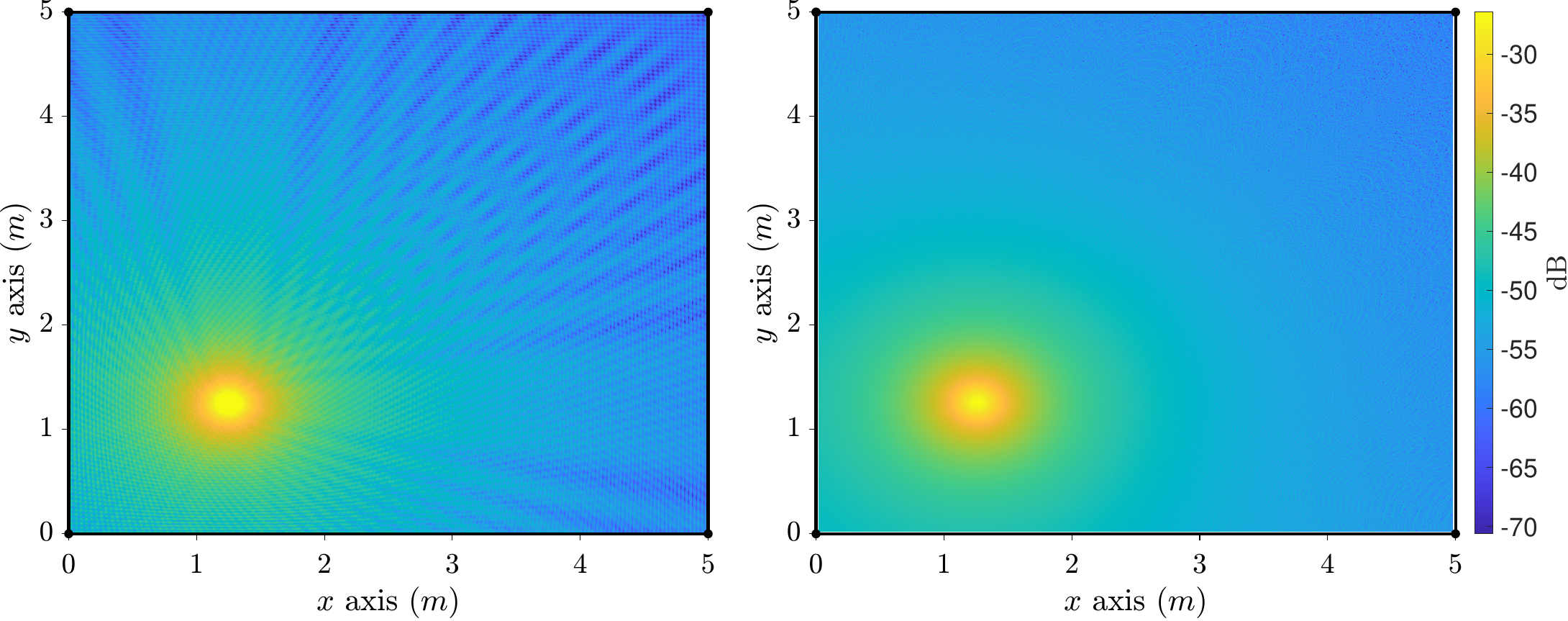}
		\caption{Layout 3, $f$ = 5 GHz}
		\label{fig:sub11111}
	\end{subfigure}
	\begin{subfigure}{0.49\linewidth}
		\centering
		\includegraphics[width=1\linewidth]{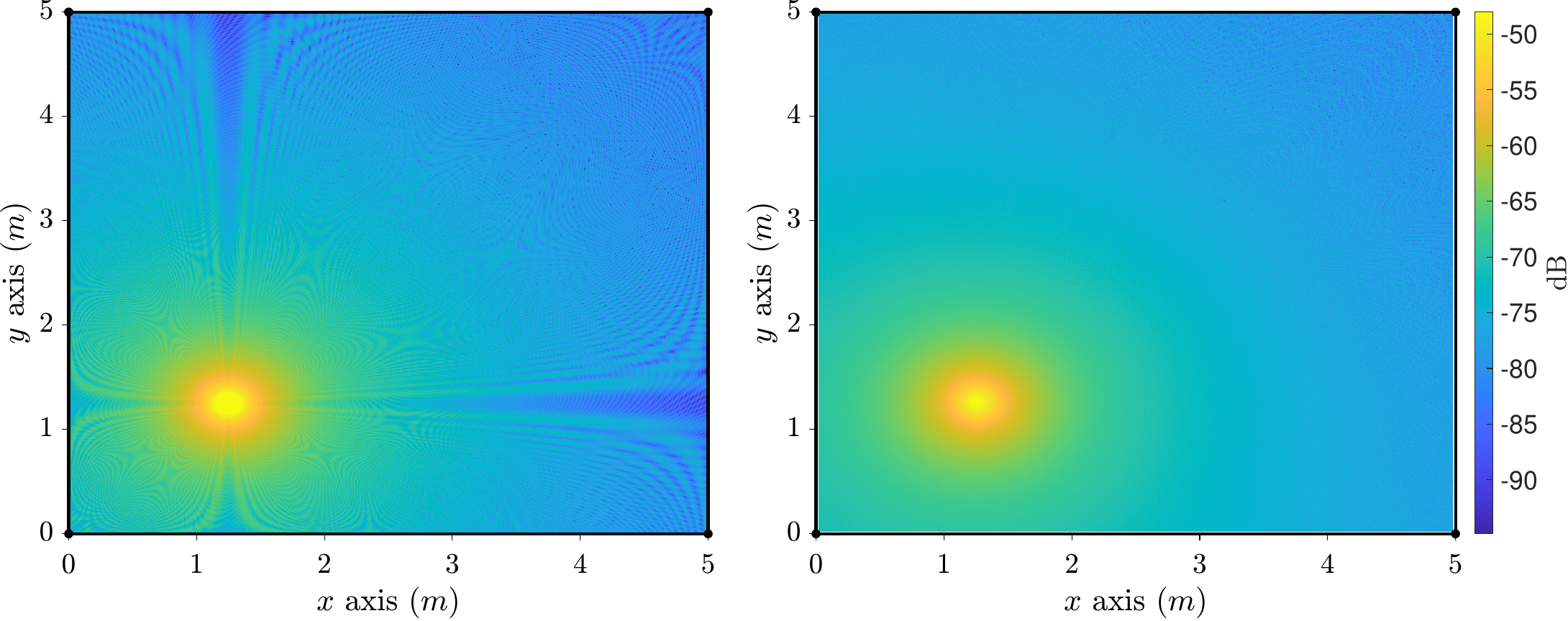}
		\caption{Layout 3, $f$ = 60 GHz}
		\label{fig:sub111111}
	\end{subfigure} 
	\caption{Comparison with Sionna, for each layout, the left one is our PL heatmap and the right one is Sionna's \cite{hoydis2022sionna} PL heatmap.}
	\label{9}
\end{figure*}

\begin{table*}[t]
	\renewcommand\arraystretch{1.4}
	\tabcolsep=0.31cm
	\centering		 
	\captionof{table}{\textsc{Performance Comparison of Proposed Channel Model with Sionna} \cite{hoydis2022sionna}.} \label{Tab1}
	\centering
	\begin{tabular}{c|cc|cc|c|cc|c|cc}
		\hline
\multirow{2}{*}{\makecell{\textbf{Layout}, \\ \textbf{Frequency}}} & \multicolumn{2}{c|}{\textbf{\begin{tabular}[c]{@{}c@{}}MAE\\ (dB)\end{tabular}}} & \multicolumn{2}{c|}{\textbf{\begin{tabular}[c]{@{}c@{}}RMSE\\ (dB)\end{tabular}}} & \textbf{\begin{tabular}[c]{@{}c@{}}Average Rate\\ (bps/Hz)\end{tabular}} & \multicolumn{2}{c|}{\textbf{\begin{tabular}[c]{@{}c@{}}Average Rate\\ (bps/Hz), Sionna\end{tabular}}} & \textbf{\begin{tabular}[c]{@{}c@{}}CPU \\ Times (s)\end{tabular}} & \multicolumn{2}{c}{\textbf{\begin{tabular}[c]{@{}c@{}} CPU Times (s),\\ Sionna\end{tabular}}} \\ \cline{2-11} 
                                 & \multicolumn{1}{c|}{\textbf{$1^\text{st}\&2^\text{nd}$}}                 & \textbf{Full}                & \multicolumn{1}{c|}{$1^\text{st}\&2^\text{nd}$}                 & \textbf{Full}                 & \textbf{Ours}                                                            & \multicolumn{1}{c|}{$1^\text{st}\&2^\text{nd}$}                           & \textbf{Full}                           & \textbf{Ours}      & \multicolumn{1}{c|}{$1^\text{st}\&2^\text{nd}$}                     & \textbf{Full}                     \\ \hline
		{1, 5GHz}      & \multicolumn{1}{c|}{1.8234} & 1.8496 & \multicolumn{1}{c|}{2.5542} & 2.5935 & 15.7479 & \multicolumn{1}{c|}{16.0936} & 16.1032 &679.8&\multicolumn{1}{c|}{2023.4} & 3609.8 \\ \hline
		{1, 60GHz}     & \multicolumn{1}{c|}{1.9835} & 2.0092 & \multicolumn{1}{c|}{2.7924} & 2.8251 & 10.7185 & \multicolumn{1}{c|}{11.0554} & 11.0659 &686.6&\multicolumn{1}{c|}{1858.4} & 3487.4 \\ \hline
		{2, 5GHz}      & \multicolumn{1}{c|}{2.2681} & 2.2498 & \multicolumn{1}{c|}{3.5297} & 3.3586 & 17.6271 & \multicolumn{1}{c|}{17.8886} & 17.9231 &575.5&\multicolumn{1}{c|}{1896.7}&3547.0 \\   \hline 
		  {2, 60GHz}     & \multicolumn{1}{c|}{2.3982} & 2.3610 & \multicolumn{1}{c|}{3.7151} & 3.4513 & 11.8190 & \multicolumn{1}{c|}{12.1181} & 12.1568 &569.6&\multicolumn{1}{c|}{1855.1}&3505.4\\  \hline
		{3, 5GHz}      & \multicolumn{1}{c|}{1.9391} & 1.9542 & \multicolumn{1}{c|}{2.6944} & 2.7179 & 22.3751 & \multicolumn{1}{c|}{22.6285} & 22.6379 &483.1&\multicolumn{1}{c|}{1647.5}&3294.2\\    \hline
		  {3, 60GHz}     & \multicolumn{1}{c|}{2.1596} & 2.1737 & \multicolumn{1}{c|}{3.0363} & 3.0576 & 15.0859 & \multicolumn{1}{c|}{15.4552} & 15.4667 &485.7&\multicolumn{1}{c|}{1646.3}&3275.5
		\\ \hline			
	\end{tabular}
\end{table*}

 	 	 \begin{figure*}[t]  
 	 		\begin{subfigure}{0.485\linewidth}
 	 			\centering
 	 			\includegraphics[width=0.825\linewidth]{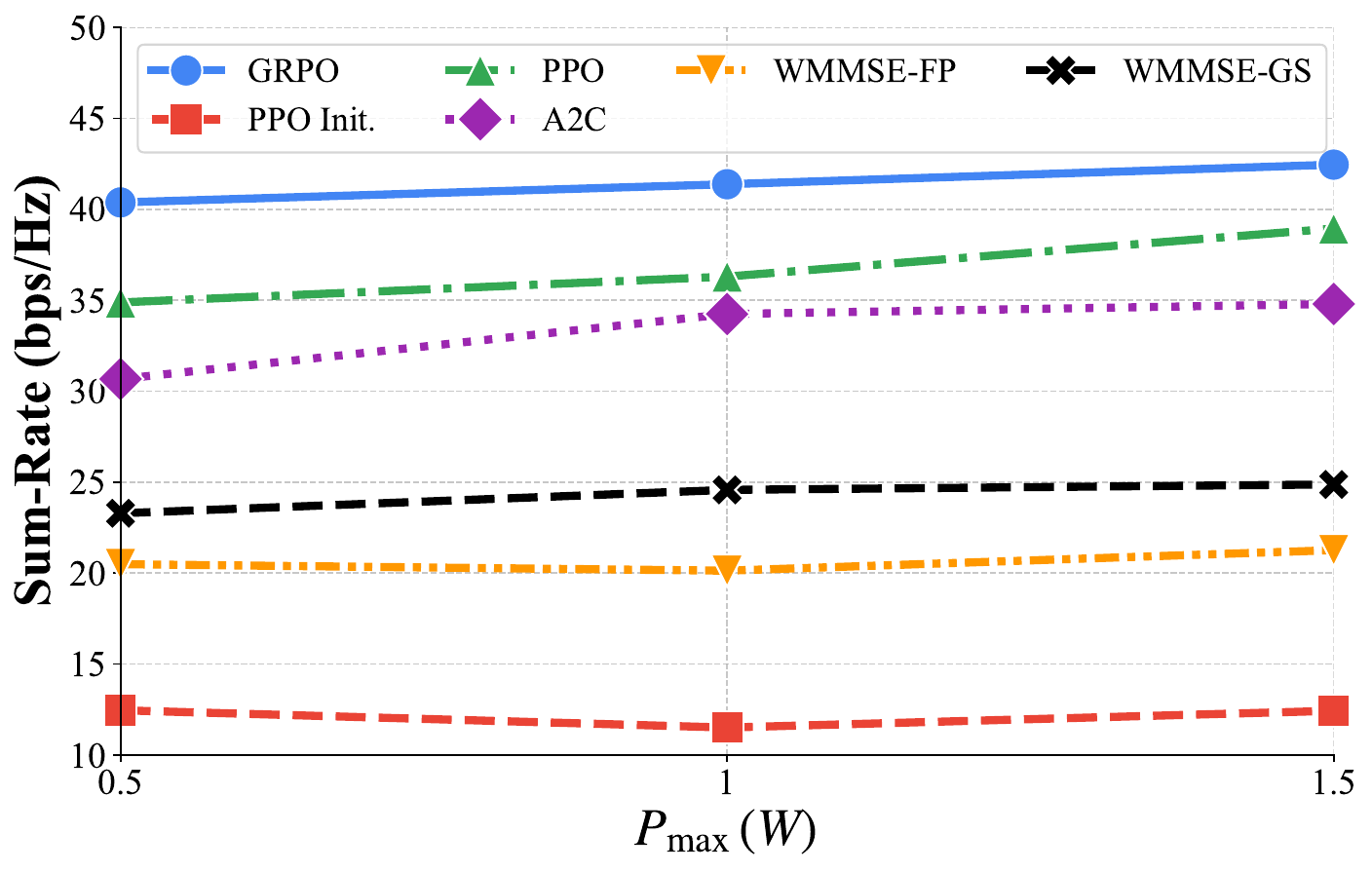}
 	 			\caption{{GRPO v.s. baselines with different $P_{\max}$.}} \label{12}

 	 		\end{subfigure}
 	 		\hfill
 	 		\begin{subfigure}{0.245\linewidth}
 	 			\centering
 	 			\includegraphics[width=0.96\linewidth]{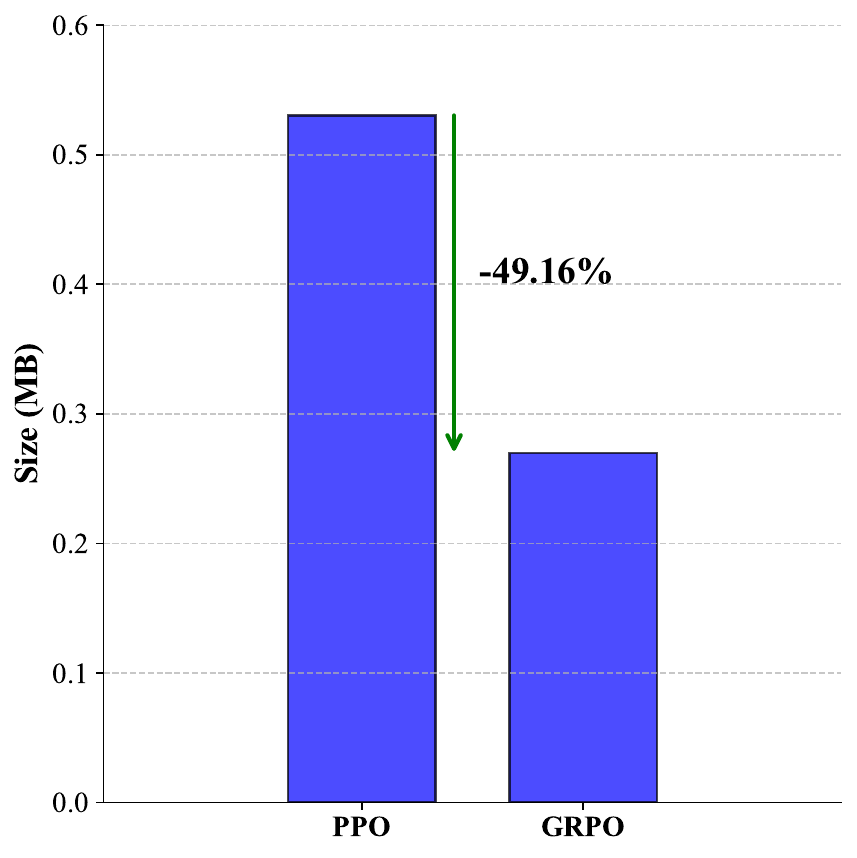}
 	 			\caption{Model size reduction.} \label{Modelsize}

 	 		\end{subfigure}   
 	 		\begin{subfigure}{0.245\linewidth}
 	 			\centering
 	 			\includegraphics[width=0.98\linewidth]{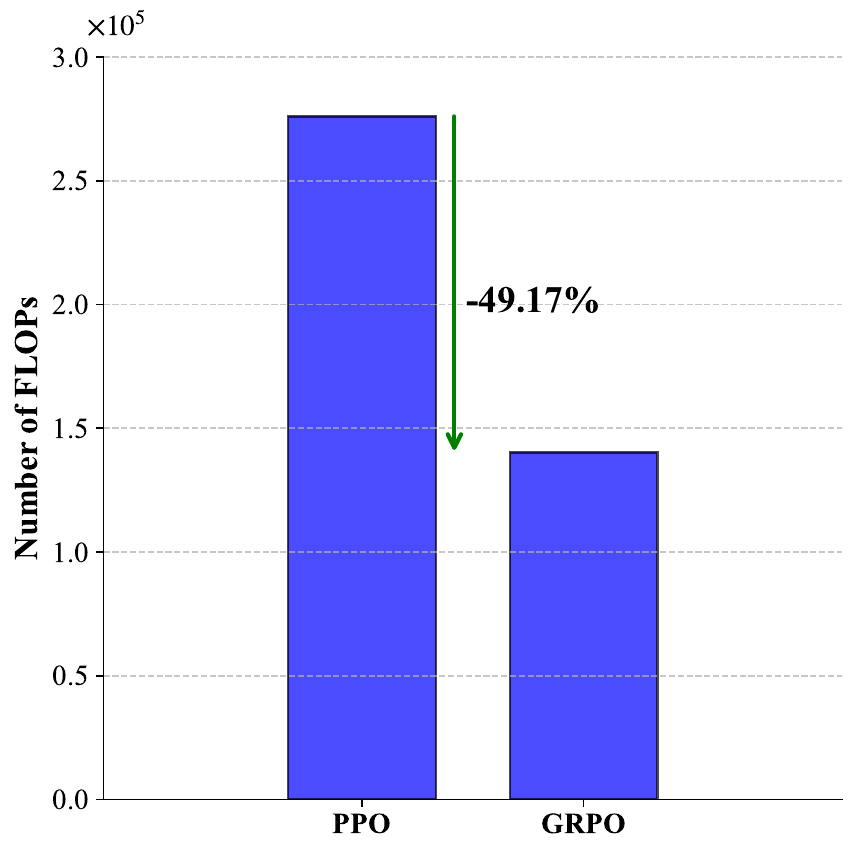}
 	 			\caption{FLOPs reduction.}	\label{FLOPs}

 	 		\end{subfigure}
 	 		\caption{{Sum-rate, model size, and FLOPs comparison of GRPO with baselines.}}  \label{100}
 	 	\end{figure*}
		
 	 	\begin{figure*}[t]
 	 		\begin{subfigure}{0.49\linewidth}
 	 			\centering
 	 			\includegraphics[width=0.825\linewidth]{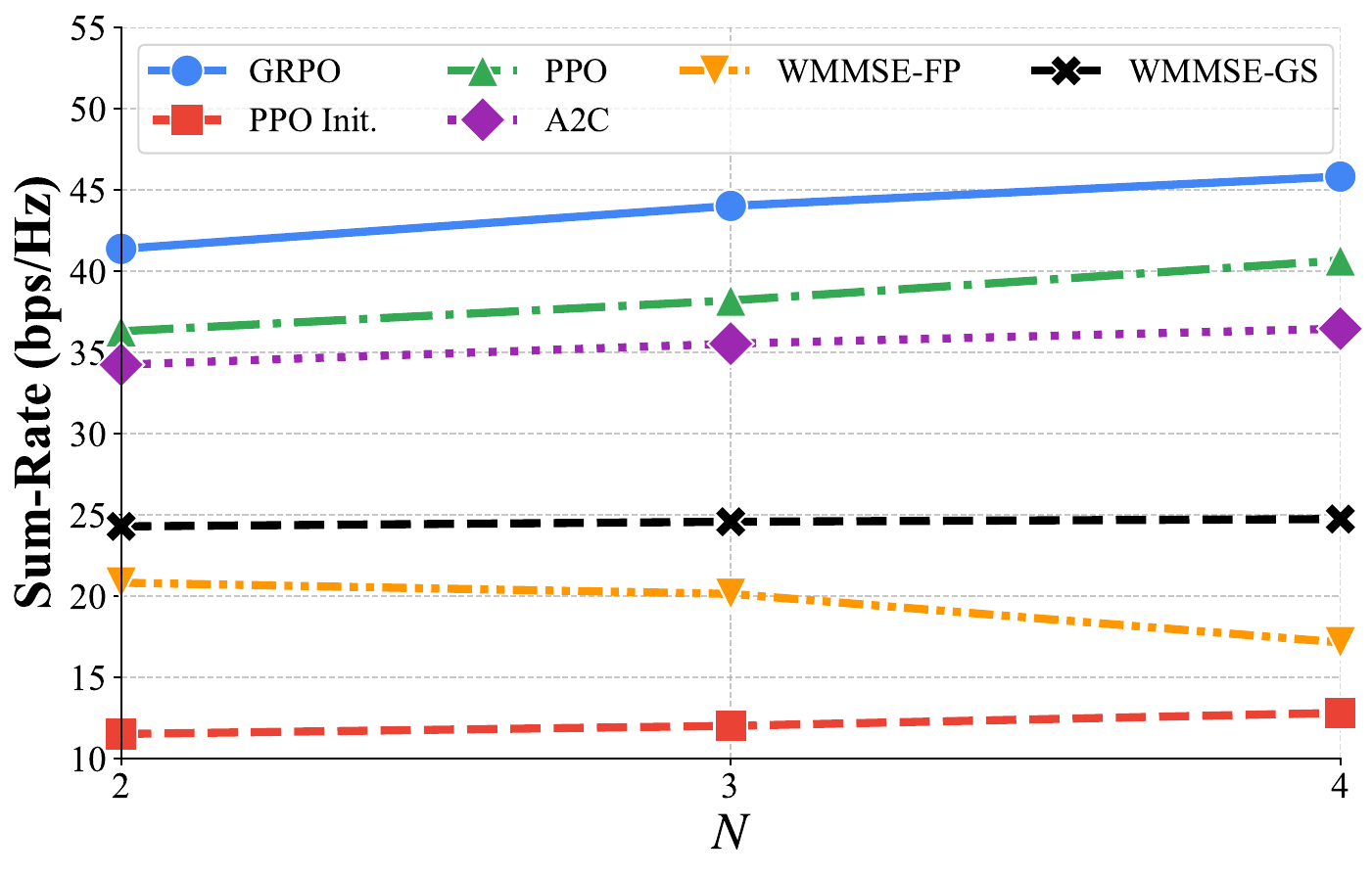}
 	 			\caption{{GRPO v.s. baselines with different $N$.}} \label{Num1}

 	 		\end{subfigure}
 	 		\hfill
 	 		\begin{subfigure}{0.49\linewidth}
 	 			\centering
 	 			\includegraphics[width=0.975\linewidth]{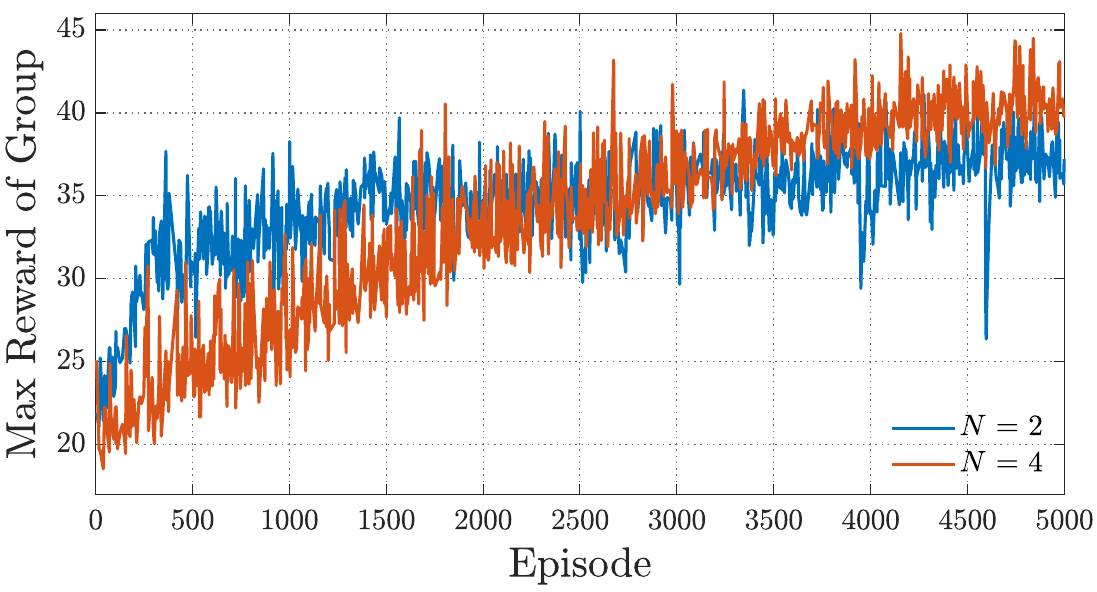}
 	 			\caption{GRPO training with $N=2$ and $N=4$.} \label{Num2}

 	 		\end{subfigure}
 	 		\caption{{Sum-rate comparison of GRPO with baselines.}}  \label{Num0}
 	 	\end{figure*}
 	 	\begin{figure*}
 	 	\begin{subfigure}{0.49\linewidth}
 		\centering
 		\includegraphics[width=0.90\linewidth]{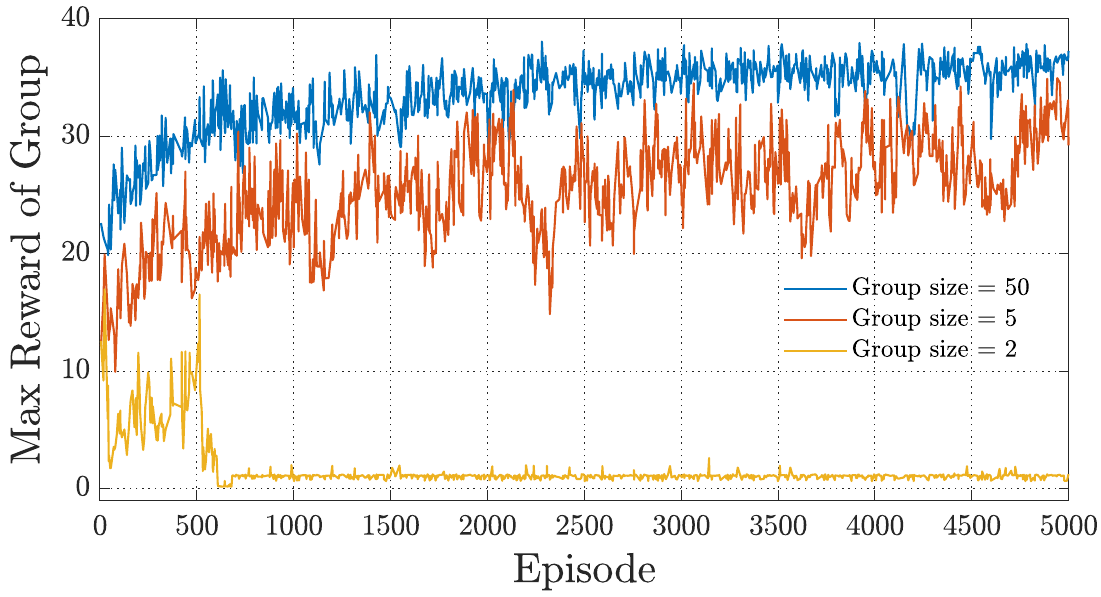}
 		\caption{GRPO with different group size when $P_{\max} = 0.5$W.} \label{Groupsize1}
 		\vspace{0.15cm}
 	\end{subfigure}
 	\hfill
 		\begin{subfigure}{0.49\linewidth}
 		\centering
 		\includegraphics[width=0.90\linewidth]{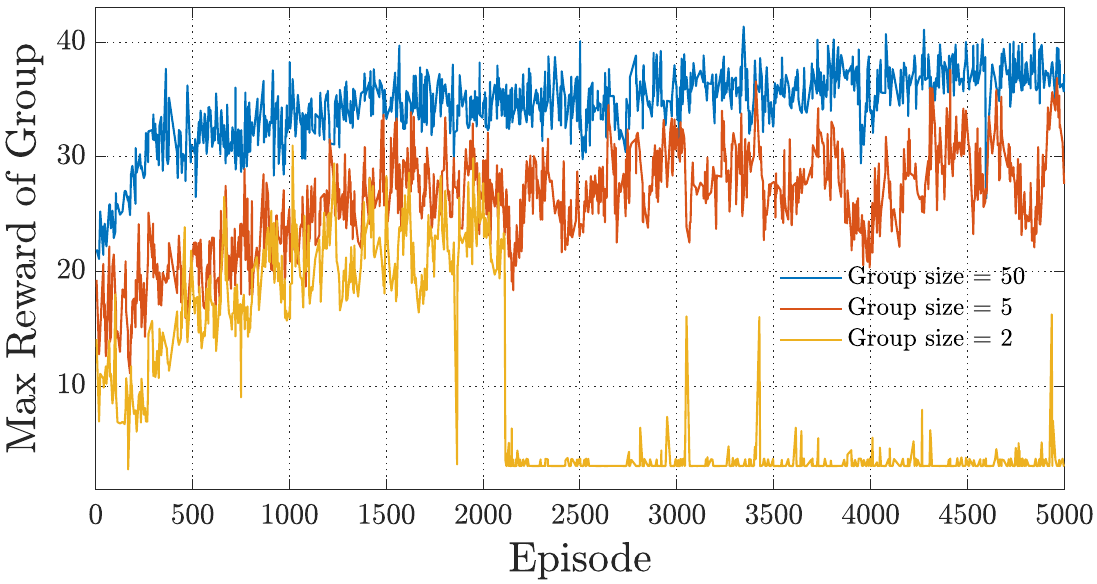}
 		\caption{GRPO with different group size when $P_{\max} = 1$W.} \label{Groupsize2}
 		\vspace{0.15cm}
 	\end{subfigure} 	
 	\hfill  
 \begin{subfigure}{0.49\linewidth}
 	\centering
 	\includegraphics[width=0.90\linewidth]{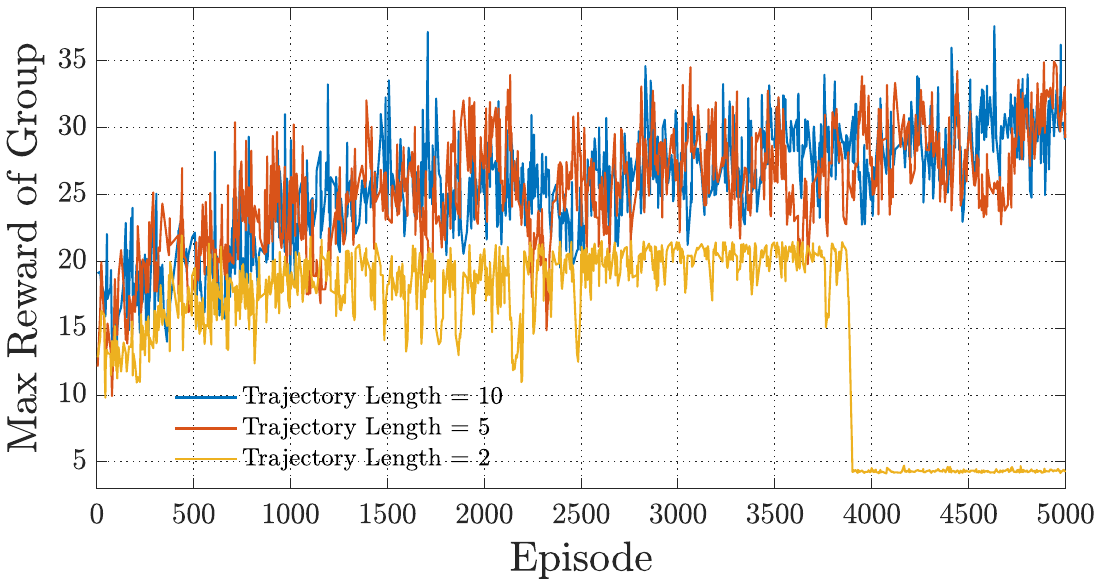}
 	\caption{GRPO with different trajectory length when $P_{\max} = 0.5$W.} \label{Length1}
 	\vspace{0.15cm}
 \end{subfigure}
 \hfill
 	\begin{subfigure}{0.49\linewidth}
 		\centering
 		\includegraphics[width=0.90\linewidth]{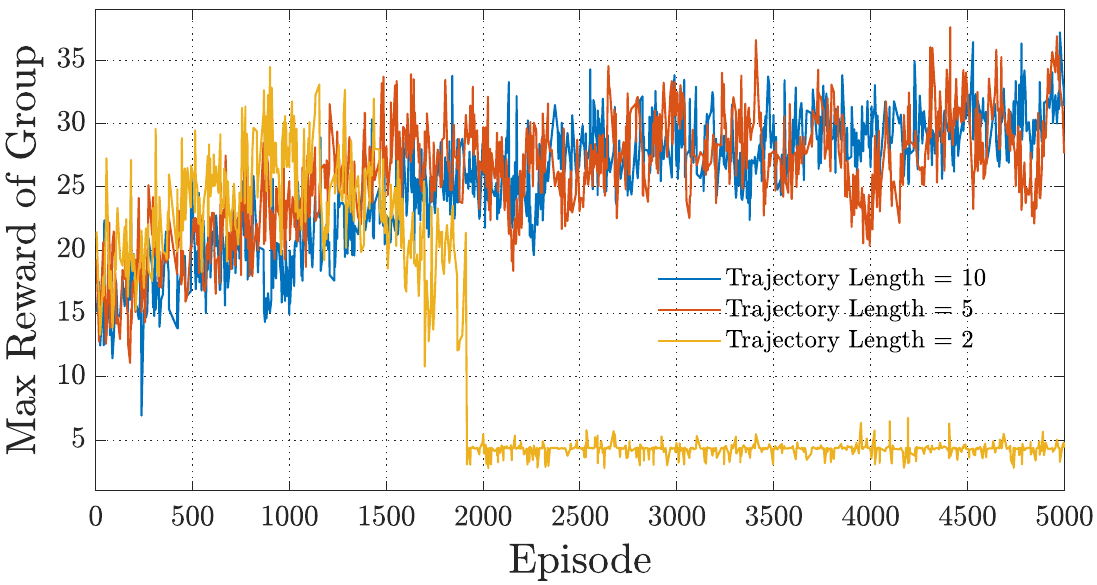}
 		\caption{GRPO with different trajectory length when $P_{\max} = 1$W.} \label{Length2}
 		\vspace{0.15cm}
 	\end{subfigure}
 	 \begin{subfigure}{0.49\linewidth}
 		\centering
 		\includegraphics[width=0.90\linewidth]{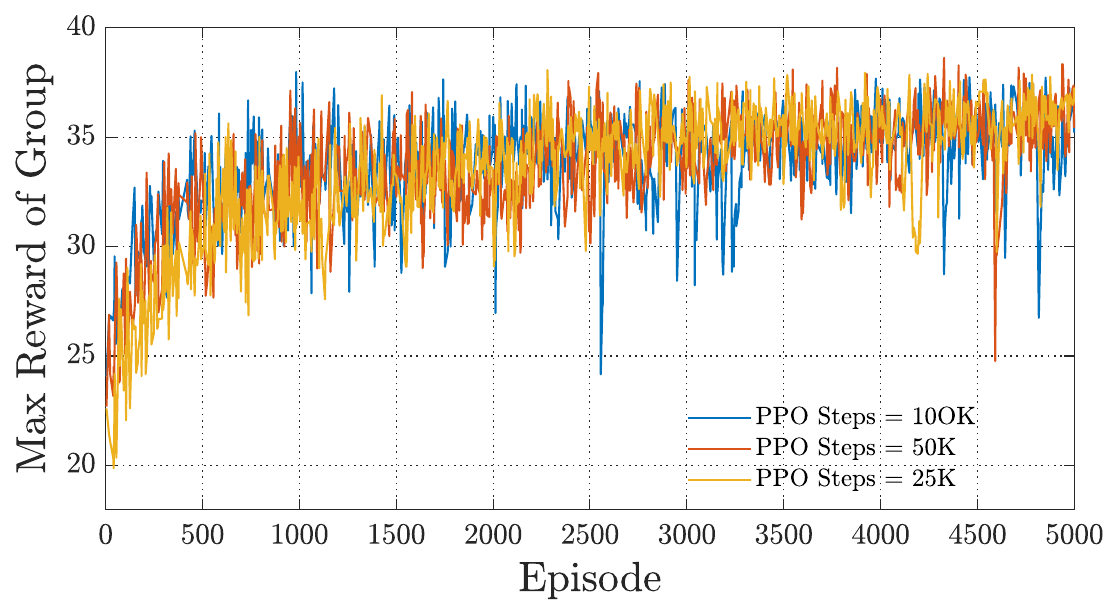}
 		\caption{GRPO with different PPO initialization steps when $P_{\max} = 0.5$W.} \label{Init1}
 	\end{subfigure}
 	\hfill
 	\begin{subfigure}{0.49\linewidth}
 		\centering
 		\includegraphics[width=0.90\linewidth]{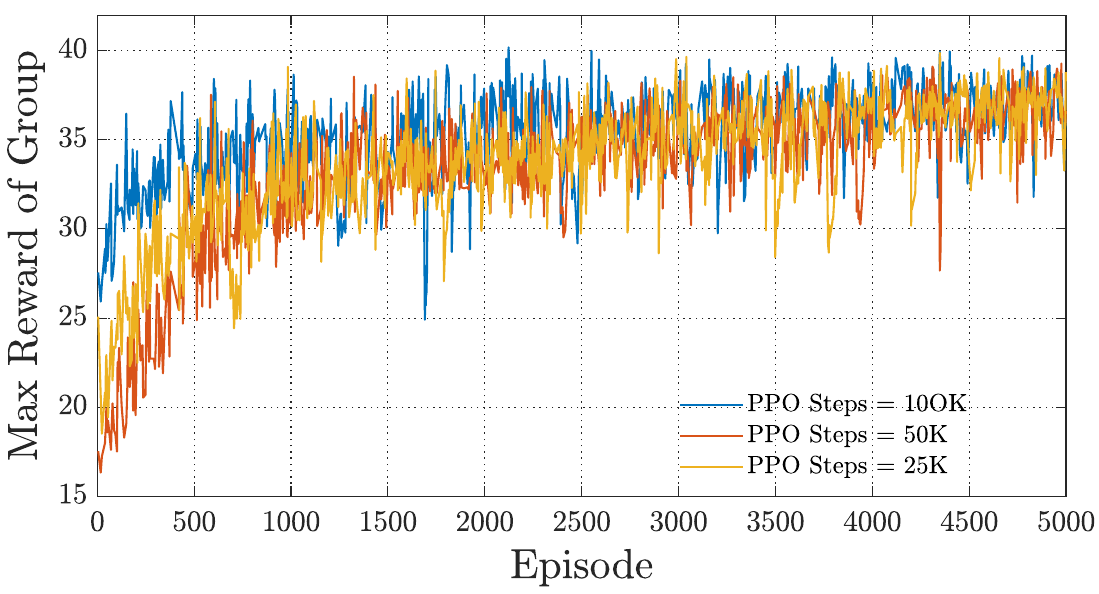}
 		\caption{GRPO with different PPO initialization steps when $P_{\max} = 1$W.} \label{Init2}
 	\end{subfigure}
 	\caption{Comparison of different group sizes, trajectory length, and PPO initialization steps for GRPO.} \label{101}
 \end{figure*}

\subsection{Performance Evaluation}

In Fig. \ref{7}, we examine the performance of the proposed closed-form solution, where Fig. \ref{7} shows different Tx and Rx placement.  Fig. \ref{7} shows that the proposed closed-form solution is exactly the same as the globally optimal solution by exhaustive search for $0.23 \pi \le \theta \le 0.4\pi$ and $0.6\pi \le \theta \le 0.82\pi$.  While, there is a slight difference between the proposed closed-form solution and the globally optimal solution by exhaustive search for $0.4\pi \le \theta \le 0.6\pi$. This can be interpreted as follows: The behavior of the rate function in $0.23\pi \le \theta \le 0.4\pi$ and $0.6\pi \le \theta \le 0.82\pi$ is primarily determined by a pronounced increasing or decreasing trend. Therefore, the proposed closed-form solution, that is, the local maximizer near the boundary, is globally optimal. In contrast, within $0.4\pi \le \theta \le 0.6\pi$, the rate function exhibits a non-monotonic behavior, increasing until $\pi/2$ and then decreasing. Thus, the proposed closed-form solution, i.e., the local maximizer near to $\pi/2$, may not be globally optimal, which is validated in Fig. \ref{7}. As for $0.4\pi \le \theta \le 0.6\pi$, Fig. \ref{7} shows that the proposed solution approaches the globally optimal solution, having a near-optimal performance.

{In Table \ref{tab:methods_scenarios}, we further examine the proposed closed-form solution in a practical layout and compare it with PPO init. baseline. Table \ref{tab:methods_scenarios} demonstrates that the proposed closed-form solution outperforms the PPO init. baseline, even after the latter's 25 K-steps training. This performance gap widens as the distance between the Tx and Rx increases. The reason for this is that NLoS rays reflected from other walls become progressively weaker compared to the two-rays, consisting of a wall reflection NLoS ray and the LoS ray, utilized by the proposed closed-form solution. Table \ref{tab:methods_scenarios} shows that our physics-informed closed-form solution could outperform general-purpose algorithms.} 

In Fig. \ref{9}, we compare the radio maps generated by the proposed layout-specific RT model with those from Sionna with full-order reflection \cite{hoydis2022sionna}, where $G_t=G_r=1$,  $\epsilon =  5.24$, TM mode, and single antenna. Fig. \ref{9} reveals a strong consensus in the spatial PL between our model and Sionna generations, except that our model exhibits more pronounced coloration in regions distant from Tx and near corners, whereas Sionna does not show this feature. This discrepancy arises from the fact that our model only accounts for first-order reflection, which cannot reach these specific regions, whereas Sionna's algorithm incorporates full-order reflection paths.

{A detailed comparison of Fig. \ref{9} is presented in Table \ref{Tab1}. The results indicate that our model achieves mean absolute error (MAE) and RMSE of approximately 2 dB and 3 dB for Sionna $1^\text{st}\&2^\text{nd}$- and full-order reflection models. Furthermore, the average rate over the radio map (calculated at a 1 cm² resolution) is comparable to Sionna $1^\text{st}\&2^\text{nd}$- and full-order reflection models. The negligible performance difference between Sionna $1^\text{st}\&2^\text{nd}$- and full-order models implies that the first and second-order reflections are dominant in these 3 scenarios. Most significantly, our model offers a substantial computational advantage, requiring only $1/3$ and $1/6$ of the time of Sionna second and full-order models, respectively.}

In Fig. \ref{100}, we evaluate the performance of the GRPO solution across different maximal transmit power.
The simulation computer platform is with NVIDIA GeForce Laptop 4090 16G and INTEL Core i9-13600. We consider a rectangular layout with corner points $\{(0, 0), (0, 5), (5, 5), (5, 0)\}$ (Unit: m). A two-antenna transmitter is located at $(x_0, 0.5\,\text{m})$, where $x_0 \in [3.5, 4.25]\,\text{m}$, and two receivers are located at $(1.25, 1.25)\,\text{m}$ and $(4.25, 3.0)\,\text{m}$, respectively. The frequency is $5\,\text{GHz}$. For GRPO, we take group size $50$ and trajectory length $5$. 
Specifically, Fig. \ref{12} highlights the superior sum-rate achieved by GRPO compared to baselines listed above, exhibiting a significant improvement over the GRPO reference policy. This result highlights GRPO's capability to transform poorly-performing policies into ones that deliver superior performance.

Furthermore, Fig.~\ref{12} shows that as maximal transmit power increases, GRPO consistently outperforms all baselines, with a notable gain at a higher maximal transmit power. This implies the effectiveness of GRPO in jointly optimizing the antenna positions, beamforming vectors, and transmit power allocation, especially when abundant transmission resources are available. Fig. \ref{Modelsize} shows a 49.2\% reduction in model size when using GRPO compared to PPO. Aligned with \cite{GRPO}, this reduction shows that GRPO can achieve better performance with a more compact neural network model. 

Fig. \ref{FLOPs} demonstrates a significant reduction in FLOPs during training when using GRPO, primarily attributed to the elimination of the value model and optimized group-based sampling. Compared to traditional PPO, GRPO achieves a substantial reduction in computational overhead, reflecting its superior efficiency in resource utilization. This efficiency gain directly translates to lower training costs, making GRPO a more practical and scalable solution for large-scale model training, particularly in resource-constrained environments.

In Fig. \ref{Num0}, we evaluate the performance of the GRPO solution across different number of transmit antennas. Unless specified otherwise, we use the same parameters and experiment environment as in Fig. \ref{100}. Here, we take group size $50$, trajectory length $5$, and $P_{\max} = 1\,$W. Fig. \ref{Num1} shows that except {WMMSE-FP}, all algorithms attain their maximum sum-rate at $4$ transmit antennas, since more transmit antennas can provide greater beamforming gain. The sum-rate of {WMMSE-FP} at $4$ transmit antennas is not the highest, because {WMMSE-FP} employs a fixed antenna position with randomized experiments, preventing it from fully exploiting the beamforming gain as the number of antennas grows. {In contrast, WMMSE-GS can exploit the beamforming gain by searching with a larger number of antennas on the grid.} 
In addition, Fig. \ref{Num2} shows that the maximal reward of group with $4$ transmit antennas eventually defeats the maximal reward of group with $2$ transmit antennas after $5000$ episodes. This outcome aligns with more beamforming gain at $4$ transmit antennas.

In Fig. \ref{101}, we investigate how group size, trajectory length, and PPO initialization steps affect the performance of GRPO. Unless specified otherwise, we use the same parameters and experiment environment as in Fig. \ref{100}. The group size and trajectory length are critical for balancing the celebrated trade-off between exploration and exploitation. In particular, a larger group size and trajectory length encourages exploration, while a smaller group size and trajectory length encourage exploitation \cite{GRPO}, since a reward is given at the last observation of a trajectory. Given trajectory length $5$, Figs. \ref{Groupsize1} and \ref{Groupsize2} show that regarding group maximal reward, the group size $50$ enjoys a moderate gain and less fluctuation compared with group size $5$, where the training with group size $50$ costs approximately $25$ hours while the training with group size $5$ costs approximately $4$ hours. This implies that increasing the group size could not lead to a significant enhancement but consuming a large amount of training time. Furthermore, Figs. \ref{Groupsize1} and \ref{Groupsize2} show that a small group size, i.e., group size $2$, leads to a collapse, since it is easily trapped. Given group size $5$, Figs. \ref{Length1} and \ref{Length2} show that regarding group maximal reward, trajectory length $10$ has a similar performance to trajectory length $5$. This implies that increasing the trajectory length may not lead to continuous performance enhancement. Figs. \ref{Length1} and \ref{Length2} show that a small trajectory length, i.e., trajectory length $2$, leads to a collapse as well.  Given group size $50$ and trajectory length $5$, Figs. \ref{Length1} and \ref{Length2} show that different PPO initialization steps lead to a similar performance. This shows that GRPO has an excellence optimization capability irregardless of PPO initialization steps. 
To summarize, the results in Fig. \ref{101} highlight the importance of tailoring the group size and trajectory length to the environmental complexity and computational constraints.

\section{Conclusion}\label{sec:conclude}
This paper investigated how FAS can enhance indoor wireless communications by tackling critical challenges in channel modeling and system optimization. Specifically, our main contributions include: 1) we developed a layout-specific ray-tracing model that achieves an $83.3$\% reduction in computational time compared to Sionna RT while only $3$ dB RMSE loss, enabling real-time multipath tracking; 2) we derived a closed-form solution for the two-ray model that preserves near-optimal performance; and 3) we proposed GRPO optimization solution for jointly optimizing antenna positions, beamforming, and power allocation. Our simulation results revealed that increasing the group size and trajectory length for GRPO does not yield prominent sum-rate improvement while consuming a large amount of training time, suggesting the importance of group size and trajectory length selection.  In addition, GRPO solution showed superior sum rate performance over the PPO, A2C, and WMMSE approaches while requiring $49.2\%$ fewer FLOPs than PPO. {In the future, our indicator function framework can be extended to incorporate second-order reflection by computing two reflection points.}

\begin{appendices}
				\section{{Angle-Dependent Distance Derivation Example}}			
		Consider the room layout illustrated in Fig. \ref{F3} as an example. {Since walls are either perpendicular to $x$-axis or $y$-axis, it can be easily observed from Fig. \ref{F3} that}
			\begin{subequations}
				\begin{eqnarray}
					&&	D^{[1]}(\mathcal{L},\mathcal{X}) = 
					\begin{cases}
						x_0, &  \qquad \text{Wall 1}, \\
						{y_1^c - y_0}, & \qquad \text{Wall 2}, 	\\
						x_5^c - x_0,  & \qquad \text{{Wall 5}}, \\
						{y_0}, 	& \qquad \text{Wall 6},
					\end{cases} \label{D1} \\
					&&		D^{[2]}(\mathcal{L},\mathcal{X}) = 	\begin{cases}
						x_1, & \qquad \text{Wall 1}, \\
						{y_1^c - y_1}, & \qquad \text{Wall 2}, 	\\
						x_5^c - x_1,  & \qquad \text{{Wall 5}}, \\
						{y_1}, 	& \qquad \text{Wall 6},
					\end{cases}  \label{D2} \\
					&&		D^{[3]}(\mathcal{L},\mathcal{X}) =  \begin{cases}
						y_1 - y_0, & \quad\,\,\,\,\,\,    \text{Wall 1}, \\
						{x_1 - x_0}, & \quad  \,\,\,\,\,\, \text{Wall 2}, 	\\
						y_1 - y_0,  & \quad  \,\,\,\,\,\,  \text{{Wall 5}}, \\
						{x_1-x_0}, 	& \quad \,\,\,\,\,\, \text{Wall 6}.
					\end{cases} \label{D3}
				\end{eqnarray}
			\end{subequations}            
	By means of substituting \( x_0 = x_1 - \frac{y_1 - y_0}{\tan \theta} \) into  \eqref{D1}-\eqref{D3}, we derive the angle-dependent distance functions \( D^{[1]}(\theta) \), \( D^{[2]}(\theta) \), and \( D^{[3]}(\theta) \). This ends the example.

			\section{Proof of Theorem 1} 	 
	 It can be seen that $D^{[1]}(\theta) = x_1 - \frac{y_1 - y_0}{\tan \theta}$, $D^{[2]}(\theta) = x_1$, and $D^{[3]}(\theta) = y_1 - y_0$.  Thus, we  have
			\begin{equation}
				d_\text{NLoS} = d_1 + d_2 = \sqrt{(y_1 - y_0)^2 + \left(2x_1 - \frac{y_1 - y_0}{\tan \theta} \right)^2},
			\end{equation}
		 $\Gamma(\theta)$ is given by \eqref{GGamma} with 
			\begin{equation}
				\alpha  = \arctan \left( \frac{2x_1}{y_1 - y_0} - \frac{1}{\tan \theta}\right),
			\end{equation} and $d_\text{LoS}$ is given in \eqref{22}.
		This leads to			
			\begin{eqnarray}
				&& \!\!\!\!\!\!\!\!\!\!\!\!\!\! h_{2}(\theta) = \frac{\sqrt{G_t G_r} \lambda}{4\pi}   \times \nonumber \\ && \!\!\!\!\!\!\!\!\!\!\!\!\!\! \left( \frac{e^{-j2\pi \frac{y_1 - y_0}{\lambda \sin \theta}}}{d_\text{LoS}} 
			 +   \frac{\Gamma e^{-j2\pi \frac{\sqrt{(y_1 - y_0)^2 + \left(2x_1 - \frac{y_1 - y_0}{\tan \theta}\right)^2}}{\lambda}}}{\sqrt{(y_1 - y_0)^2 + \left(2x_1 - \frac{y_1 - y_0}{\tan \theta}\right)^2}}  \right). \label{RG5}
			\end{eqnarray}						
	  Since $\textsc{SNR}_{2}(\theta) = \frac{|h_{2}(\theta)|^2}{\sigma^2}$, we have 
			\begin{eqnarray}
				&&	\!\!\!\!\!\!\!\!\!\!\!\!\! \textsc{SNR}_{2}(\theta) =   \frac{G_tG_r\lambda^2}{\sigma^2\left(4\pi\right)^2}  \left( \left( \frac{\sin \theta \cos (2\pi \frac{y_1-y_0}{\lambda \sin \theta})}{y_1 -y_0}     \right.\right.  \nonumber \\
				&& \!\!\!\!\!\!\!\!\!\!\!\!\!  \qquad \qquad \qquad +  \left. \frac{\Gamma \cos (2\pi \frac{\sqrt{(y_1 - y_0)^2 + (2x_1 - \frac{y_1 - y_0}{\tan \theta})^2}}{\lambda})}{\sqrt{(y_1 - y_0)^2 + (2x_1 - \frac{y_1 - y_0}{\tan \theta})^2}} \right)^2     \nonumber \\
				&& \!\!\!\!\!\!\!\!\!\!\!\!\!   \qquad \quad \quad  +    \left(   \frac{\sin \theta\sin (2\pi \frac{y_1-y_0}{\lambda \sin \theta})}{y_1 - y_0} \right. \nonumber \\
				&& \!\!\!\!\!\!\!\!\!\!\!\!\!   \qquad \quad \quad \left. \left. +   \frac{\Gamma \sin (2\pi \frac{\sqrt{(y_1 - y_0)^2 + (2x_1 - \frac{y_1 - y_0}{\tan \theta})^2}}{\lambda})}{\sqrt{(y_1 - y_0)^2 + (2x_1 - \frac{y_1 - y_0}{\tan \theta})^2}} \right)^2 \right).  
				\label{RG7}
			\end{eqnarray}	 
			By expanding \eqref{RG7} and applying the trigonometric identity $\cos^2(\cdot)  + \sin^2(\cdot)  = 1$, we can prove Theorem 1.

		 	\section{Proof of Proposition 1}
		 First of all, note that we aim at finding $\widetilde{\theta}$ satisfying  
		 \begin{eqnarray}				 
		 	&&\!\!\!\!\!\!\!\!\!\! \cos \left( 
		 	\frac{2 \pi 		 		
		 	}{\lambda} \left( \sqrt{(y_1 - y_0)^2 + 
		 			\left(2x_1 - \frac{y_1 - y_0}{\tan \widetilde{\theta}}\right)^2} 
		 		- \frac{y_1 - y_0}{\sin \widetilde{\theta}}
		 	\right)\right) \nonumber \\
		 	&&\!\!\!\!\!\!\!\!\!\! = 
		 	\begin{cases}
		 		1, & 0 < \Gamma, \\
		 		-1, & \Gamma < 0.
		 	\end{cases}				 
		 	\label{Ap1}
		 \end{eqnarray}				
		 We define
		 \begin{equation}
		\!\!\!\!\!\! 	n = \begin{cases}
		 		2\lfloor \frac{\sqrt{(y_1 - y_0)^2 + (2x_1 - \frac{y_1 - y_0}{\tan \theta_r})^2} - \frac{y_1 - y_0}{\sin \theta_r}}{\lambda} \rfloor, & 0 < \Gamma, \\
		 		2\lfloor \frac{\sqrt{(y_1 - y_0)^2 + (2x_1 - \frac{y_1 - y_0}{\tan \theta_r})^2} - \frac{y_1 - y_0}{\sin \theta_r}}{\lambda} \rceil - 1,	& \Gamma < 0.
		 	\end{cases}
		 \end{equation}
		 Next, we aim to find $\widetilde{\theta}$ that satisfies
		 \begin{equation}
		 	\sqrt{(y_1 - y_0)^2 + (2x_1 - \frac{y_1 - y_0}{\tan \widetilde{\theta}})^2} - \frac{y_1 - y_0}{\sin \widetilde{\theta}} = \frac{\lambda n}{2}.
		 \end{equation}
		 Squaring both sides and multiplying by \( \sin^2 \widetilde{\theta} \), we obtain
		 \begin{eqnarray}
		 	(4x_1^2 - \frac{n^2\lambda^2}{4}) \sin^2 \widetilde{\theta} - 4x_1(y_1 - y_0) \cos \widetilde{\theta} \sin \widetilde{\theta} = \nonumber \\
		 	n \lambda (y_1 - y_0) \sin \widetilde{\theta}.
		 \end{eqnarray}
		 Dividing both sides by \( \sin \widetilde{\theta} \), we have
		 \begin{equation}
		 	(4x_1^2 - \frac{n^2\lambda^2}{4}) \sin \widetilde{\theta} - 4x_1(y_1 - y_0) \cos \widetilde{\theta} = n \lambda (y_1 - y_0). \label{A1}
		 \end{equation}		 
		 Because $ \cos \phi = \frac{4x_1^2 - \frac{n^2\lambda^2}{4}}{\sqrt{(4x_1^2 - \frac{n^2\lambda^2}{4})^2 + (4x_1(y_1 - y_0))^2}} $ and $ \sin \phi = \frac{-4x_1(y_1 - y_0)}{\sqrt{(4x_1^2 - \frac{n^2\lambda^2}{4})^2 + (4x_1(y_1 - y_0))^2}} $, we have 
         \begin{equation}
              \phi = \arctan \left( \frac{-4x_1(y_1 - y_0)}{4x_1^2 - \frac{n^2\lambda^2}{4}} \right). 
         \end{equation}
    Hence, we can now re-write \eqref{A1} into
		 \begin{eqnarray}
		 	\sqrt{(4x_1^2 - \frac{n^2\lambda^2}{4})^2 + (4x_1(y_1 - y_0))^2} \sin(\widetilde{\theta} + \phi) = \nonumber \\
		 	n \lambda (y_1 - y_0). \label{A2}
		 \end{eqnarray}				
		 
		 Finally, we invoke the \( \arcsin \) function to derive \( \widetilde{\theta} \) from \eqref{A2} and let $f(n) = \widetilde{\theta} $. This completes the proof.
\end{appendices}

\bibliographystyle{IEEEtran}

\end{document}